\documentclass[useAMS,usenatbib]{mn2e}
\usepackage{graphicx}
\usepackage{amsmath,amssymb}
\usepackage[dvipdfm]{color}
\newcommand{\del}{\partial}

\newcommand{\THE}{{\boldsymbol{\theta}}}
\newcommand{\ALP}{{\boldsymbol{\alpha}}}
\newcommand{\BETA}{{\boldsymbol{\beta}}}
\newcommand{\largec}{{\boldsymbol{C}}}
\newcommand{\K}{{\boldsymbol{k}}}
\newcommand{\x}{{\boldsymbol{x}}}

\newcommand{\f}{\frac}

\newcommand{\BF}{\begin{figure}\begin{center}}
\newcommand{\EF}{\end{center}\end{figure}}
\newcommand{\BE}{\begin{equation}}
\newcommand{\EE}{\end{equation}}
\newcommand{\BEA}{\begin{eqnarray}}
\newcommand{\EEA}{\end{eqnarray}}

\newcommand{\ms}{M_{\odot}}

\newcommand{\bvec}[1]{\mbox{\boldmath $#1$}}
\begin{document}
\title{Weak lensing by intergalactic mini-structures 
in quadruple lens systems: Simulation and Detection }
\author[Ryuichi Takahashi and Kaiki Taro Inoue]
{Ryuichi Takahashi$^1$ and 
Kaiki Taro Inoue$^2$\thanks{E-mail:kinoue@phys.kindai.ac.jp}
\\
$^{1}$Faculty of Science and Technology, Hirosaki University, 3
Bunkyo-cho, 
Hirosaki, Aomori 036-8561, Japan
\\
$^{2}$Department of Science and Engineering, 
Kinki University, Higashi-Osaka, 577-8502, Japan  }


\date{\today}

\pagerange{\pageref{firstpage}--\pageref{lastpage}} \pubyear{0000}

\maketitle

\label{firstpage}
\begin{abstract}
We investigate the weak lensing effects of line-of-sight structures 
on quadruple images in quasar-galaxy strong
 lens systems based on $N$-body and ray-tracing simulations that can
 resolve halos with a mass of $\sim 10^5\,\ms$.
The intervening halos and voids disturb the magnification ratios
of lensed images as well as their relative positions due to lensing.  
The magnification ratios typically change by $O(10\%)$ 
when the shifts of relative angular positions of lensed images 
are constrained to $<0.004''$. The constrained amplitudes of projected density 
perturbations due to line-of-sight structures are 
$O(10^8)\,\ms/\textrm{arcsec}^2$. These results are consistent with our new 
analytical estimate based on the two-point correlation of density 
fluctuations. The observed mid-infrared (MIR) flux ratios for 
6 quasar-galaxy lens systems with quadruple images
agree well with the numerically estimated values without 
taking into account of subhalos 
residing in the lensing galaxies. We find 
that the constrained mean amplitudes of projected density perturbations
in the line-of-sight are negative, which suggests that the fluxes of lensed
images are perturbed mainly by minivoids and minihalos in underdense regions.
We derive a new fitting formula for estimating the 
probability distribution function of magnification perturbation. 
We also find that the mean amplitude of magnification perturbation
 roughly equals the standard deviation regardless of the model parameters. 
\end{abstract}

\section{Introduction}
In recent years, much attention has been paid to modeling of 
strong quasar-galaxy lensing systems 
with quadruple images in particular to clarifying 
the clustering property of dark matter at mass scales
of $\lesssim 1\,h^{-1} \textrm{Mpc}$ \citep{metcalf2001,chiba2002}.
In fact, the flux ratios in some quadruply lensed
quasars disagree with the prediction of best-fit lens models with a potential 
whose fluctuation scale is larger than the separation between the
lensed images. Such a discrepancy called the ``anomalous flux ratio''
has been considered as an imprint of substructure inside a lensing galaxy
\citep{mao1998,metcalf2001,metcalf2004,chiba2005,sugai2007,mckean2007,
more2009,minezaki2009,macleod2009}.

However, recent studies based on high resolution simulations 
suggested that the predicted substructure population is too low 
to explain the observed anomalous flux ratios
\citep{maccio2006,amara2006,
xu2009,xu2010,chen2009,chen2011}. 
Although detailed modeling of gravitational potential of
the lens on scales comparable to or larger than the distance between the
lensed images might also improve the fit \citep{wong2011}, the origin of the 
anomalous flux ratios in some quadruple image systems such as B1422+231
and MG0414+0534 has been veiled in mystery \citep{chiba2005, minezaki2009}. 

In addition to substructures in lensing galaxy, any intergalactic structures
 along the entire line-of-sight from the source
to the observer can also perturb the flux ratios \citep{chen2003, metcalf2005a, miranda2007, xu2012}.  
Recently, taking into account of astrometric shifts,  
\citet{it2012} (hereinafter referred to as ``IT12'') has found that 
the observed anomaly in B1422+231 and MG0414+0534 can be explained by
the line-of-sight structures without taking into account of
subhalos associated with the lensing galaxies. 
In fact, it turned out that the observed quadruple lens systems with 
high redshift sources tend to exhibit more anomalous flux ratios than
those with low redshift sources. As the 
amplitudes of convergence perturbation due to the line-of-sight structures 
increase with the source redshift, such a feature 
strongly supports a hypothesis that the
observed flux ratio anomalies are caused by intervening structures
rather than substructures associated with the primary
lens.  

However, IT12 used only 2-point correlation
function of astrometric shifts and convergence for estimating 
the mean perturbation of the flux ratios. In order to evaluate the
probability distribution of the perturbation, one needs to incorporate
the effects of higher order correlation functions as well.  
Furthermore, it is of great importance to determine the 
role of underdense regions (voids). On the one hand, 
the effect of each perturbation by an underdense region (void) 
is expected to be smaller than that by an 
overdense region (halo) since the maximum amplitude of the former 
is much smaller than that of the latter. 
On the other hand, the probability of perturbation by underdense regions 
is much larger than that by overdense regions 
since the total volume of underdense regions is much 
larger. Thus, the overall effect is non-trivial. 
To probe such non-linear effects, 
one needs to perform ray-tracing simulations using particle
distributions obtained from $N$-body simulations.

In this paper, we numerically explore the non-linear effects 
in weak lensing due to line-of-sight structures in quasar-galaxy 
quadruple lensing systems. In what follows, we assume a 
cosmology with a matter density $\Omega_m=0.272$, a baryon density 
$\Omega_b=0.046$, a cosmological constant $\Omega_\Lambda=0.728$,
the Hubble constant $H_0=70, \textrm{km}/\textrm{s}/\textrm{Mpc}$,
the spectrum index $n_s=0.97$, and the root-mean-square (rms) 
amplitude of matter fluctuations at $8 h^{-1}\, \textrm{Mpc}$, 
$\sigma_8=0.81$, which are obtained from the observed 
CMB (WMAP 7yr result, \citep{jarosik2011}), the baryon 
acoustic oscillations (Percival et~al. 2010), and $H_0$ \citep{riess2009}.

\section{Weak lensing effects of line-of-sight structures on
 magnifications of multiple images}

\begin{figure}
\vspace*{1.0cm}
\includegraphics[width=85mm]{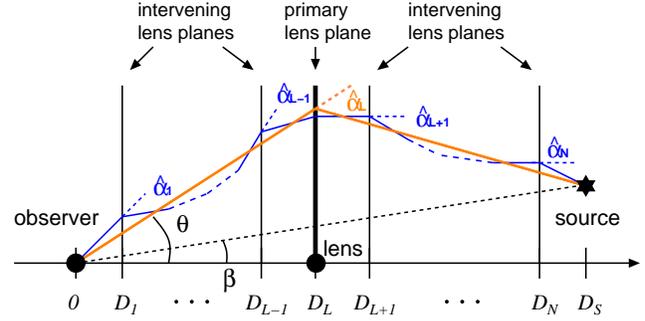}
\caption{
A configuration of multiple lens system. In the horizontal axis,
$D_{j}$ with $j=1,2,..,N$ is the angular diameter distance from the observer
 to the j-th lens plane, and here $j=L$ corresponds to the primary lens.
$\theta$ and $\beta$ are the angular positions of the image and the source.  
The blue line shows the light-ray path in the multiple scattering.
The light-ray is deflected at each lens plane by the angle $\hat{\alpha}_j$
 on the j-th lens plane.
The orange lines are the unperturbed
 light-ray paths in which we ignore
 the deflection by the intervening lens planes.
}
\label{fig_lens_planes2}
\vspace*{0.5cm}
\end{figure}

In this section, we introduce the weak lensing effects due to 
intervening halos and voids in quasar-galaxy lens systems.
First, we discuss simple primary lens systems and then 
we treat the lensing effects of line-of-sight structures as 
perturbations to the primary lens.
Our objective is to derive the magnification perturbation of each lensed
image induced by intervening structures. 

When a light ray from a distant quasar passes in the neighborhood of 
a massive lensing galaxy (primary lens), multiple images of the 
quasar are formed due to deflection of light.
The source position $\BETA$ and the image position $\THE$ are related via
 the lens equation (e.g. Schneider et al. 1992),
\BE
 \BETA = \THE - \frac{D_{LS}}{D_S} \hat{\ALP}_L (\THE),
\label{lens-eq}
\EE
where $\hat{\ALP}_L$ is the deflection angle by the primary lens,
 and $D_{LS} (D_S)$ is the angular-diameter distance from the
 lens (observer) to the source.
The image positions are obtained as solutions of the lens equation
 (\ref{lens-eq}). In what follows, we use a suffix i for labeling each image
 (${\rm i}=1,2,...$).
The image deformation is characterized by the Jacobian matrix,
 $A_L = \partial \BETA / \partial \THE$. For the i-th image, the
 matrix is
\BE
 A_{L, {\rm i}} = \begin{pmatrix} 
     1-\kappa_{\rm i}-\gamma_{{\rm i} 1} & -\gamma_{{\rm i} 2} \\
     -\gamma_{{\rm i} 2} & 1-\kappa_{\rm i}+\gamma_{{\rm i} 1} \\
                  \end{pmatrix},
\label{jacobi-mat}
\EE
where $\kappa_{\rm i}$ and $\gamma_{{\rm i} 1,2}$ are the convergence 
 and the shear of the i-th image whose angular sizes are sufficiently small.  
The magnification is $\mu_{\rm i}= 1/\det(A_{L,{\rm i}})$
 $= 1/[(1-\kappa_{\rm i})^2- \gamma_{\rm i}^2]$ where
 $\gamma_{\rm i}^2=\gamma_{{\rm i} 1}^2+\gamma_{{\rm i} 2}^2$.
The images are classified into three types: minimum, saddle, and maximum
 according to the curvature tensor of the arrival
 time surface at a lensed image 
(or the Fermat potential; e.g. Schneider et al. 1992).
 Type I is a minimum image such that
 $\gamma < 1-\kappa \leq 1$, type II is a saddle image, 
 $(1-\kappa)^2 < \gamma^2$, and type III is a maximum image, $(1-\kappa)^2 > 
 \gamma^2$ and $\kappa > 1$.

Next, we consider multiple lens planes between the observer and the
 source. The configuration of multiple lensing 
is shown in Fig. \ref{fig_lens_planes2}. 
$N$ lens planes are located at the angular-diameter distance
 $D_j$ ($j=1,2,...,N$), where $j=L$ corresponds to the primary lens while
the others ($j \neq L$) correspond to the intervening lenses.
The blue line indicates the light-ray path in the multiple scattering.
In multiple lens planes, the lens equation is given by,
\BE
 \BETA = \THE_1 - \frac{D_{LS}}{D_S} \hat{\ALP}_L (\THE_L) -
   \sum_{j \neq L} \frac{D_{jS}}{D_S} \hat{\ALP}_j (\THE_j)
\label{lens-eq-multi}
\EE
where $\THE_j$ and $\hat{\ALP}_j$ denote the light-ray position and the
 deflection angle on the j-th lens plane, respectively and
 $D_{jS}$ is the angular-diameter distance from the j-th lens plane
 to the source.
In the following, we assume that the deflection angles caused by intervening
 structures are much smaller than that by the primary lens. 
Then we can ignore the last term in equation (\ref{lens-eq-multi}).
This assumption corresponds to the Born approximation.
The orange lines in Fig. \ref{fig_lens_planes2} are unperturbed light-ray
 paths under the assumption of the Born approximation.
From equation (\ref{lens-eq}), the
 position of an unperturbed light-ray 
on the j-th plane is given by,
\BEA
  \THE_j &=& \THE_1,  ~~~{\rm for}~  j \leq L   \nonumber \\
             &=& \THE_1 - \frac{D_{Lj}}{D_j} \hat{\ALP}_L (\THE_1), 
             ~~~{\rm for}~  j > L
\EEA
where $D_{Lj}$ is the angular diameter distance 
 between the primary lens plane and the j-th lens plane. 
 From equation (\ref{lens-eq-multi}), the magnification matrix, $A=\partial
 \BETA/\partial \THE_1$, is given by
\BE
 A = A_L + \delta A, 
  ~~\delta A = - \sum_{j \neq L} \frac{D_{jS}}{D_S}
  \frac{\partial \hat{\ALP}_j (\THE_j)}{\partial \THE_1}.
\EE
Note that the derivative in $\delta A$ is calculated along the
 unperturbed path.
Then, the magnification matrix for the i-th image is,
\BEA
 A_{\rm i} = A_{L, {\rm i}} + \delta A_{\rm i}
    = \begin{pmatrix} 
     1-\kappa_{\rm i}-\gamma_{{\rm i} 1} & -\gamma_{{\rm i} 2} \\
     -\gamma_{{\rm i} 2} & 1-\kappa_{\rm i}+\gamma_{{\rm i} 1} \\
                  \end{pmatrix} \nonumber \\
     + \begin{pmatrix} 
     -\delta \kappa_{\rm i} - \delta \gamma_{{\rm i} 1}
        & -\delta \gamma_{{\rm i} 2} \\
     -\delta \gamma_{{\rm i} 2} & - \delta \kappa_{\rm i}
                    + \delta \gamma_{{\rm i} 1} \\
                  \end{pmatrix}.
\EEA
The second matrix is the correction term due to intervening structures.
The total magnification is $\mu_{\rm i} + \delta \mu_{\rm i}
 =1/\det(A_{L, \rm i}+\delta A_{\rm i})$.
The fractional difference of the magnification for the i-th image
\BEA
 \delta^{\mu}_{\rm i} \equiv \frac{\delta \mu_{\rm i}}{\mu_{\rm i}},
\label{relative_mu}
\EEA
is approximately given by
\BEA
 \delta^{\mu}_{\rm i} \simeq 2 \frac{\left( 1-\kappa_{\rm i} \right)
 \delta \kappa_{\rm i} + \gamma_{1 {\rm i}} \delta \gamma_{1 {\rm i}} +
 \gamma_{2 {\rm i}} \delta \gamma_{2 {\rm i}}}
 {\left( 1-\kappa_{\rm i}^2 \right)^2 - \gamma_{\rm i}^2},
\label{eta_approx}
\EEA
up to first order in $\delta \kappa, \delta
 \gamma_{1,2}$ and $\delta \mu$. We numerically evaluate the 
perturbations using the ray-tracing simulation. 

We use a statistic $\eta$ introduced by IT12 to measure the
magnification perturbation of lensed images in quasar-galaxy lens systems:
\BE
  \eta \equiv \biggl[\frac{1}{2 N_{\rm pair}} \sum_{{\rm i} \neq {\rm j}} \left[
 \delta^\mu_{\rm i} ({\rm minimum}) - \delta^\mu_{\rm j} ({\rm saddle})
 \right]^2 \biggr]^{1/2}
\label{eta_def}
\EE
where $\delta^\mu({\rm minimum})$ and $\delta^\mu({\rm saddle})$ are
 the magnification perturbations given
 in equation (\ref{relative_mu}) corresponding to the minimum and saddle images.
Roughlky speaking, $\eta$ is equal to the magnification contrast
averaged over the lensed images if correlation of flux perturbation 
between the lensed images are negligible. The summation is performed 
over all the pairs of the minimum and the saddle images,
 and $N_{\rm pair}$ is the total number of the pairs.
For three images of two minima A, C, and a saddle B, we have
\BE
 \eta = \frac{1}{2} \left[ \left( \delta^\mu_{\rm A} - \delta^\mu_{\rm B}
  \right)^2 + \left( \delta^\mu_{\rm C} - \delta^\mu_{\rm B} \right)^2
 \right]^{1/2}.
\EE
Similarly, for four images of two minima A, C and two saddles B, D, we have 
\BEA 
~~~~~~~~~~~~~~~\eta &=& \frac{1}{\sqrt{8}} \biggl[ \left( \delta^\mu_{\rm A}
 - \delta^\mu_{\rm B}
  \right)^2 + \left( \delta^\mu_{\rm C} - \delta^\mu_{\rm B} \right)^2
\nonumber
\\ 
&+& \left( \delta^\mu_{\rm A} - \delta^\mu_{\rm D} \right)^2
  + \left( \delta^\mu_{\rm C} - \delta^\mu_{\rm D} \right)^2 \biggr ]^{1/2}.
\EEA
To measure $\eta$, we first need to estimate 
the unperturbed fluxes $f^0_{\rm i}$ based on the best-fit lens model 
(see sec.\ref{5.1} for detail) and observed fluxes $f_{\rm i}$ for each
i-th image. Therefore, $\eta$ is a model dependent quantity. 
Then, we can estimate the magnification contrast 
$\delta^\mu_{\rm i}
 =(f_{\rm i}-f^0_{\rm i})/f^0_{\rm i}$ and also $\eta$ using the above 
 equations.

\section{Ray-tracing simulations}
\label{sec3}

\begin{figure}
\vspace*{1.0cm}
\includegraphics[width=85mm]{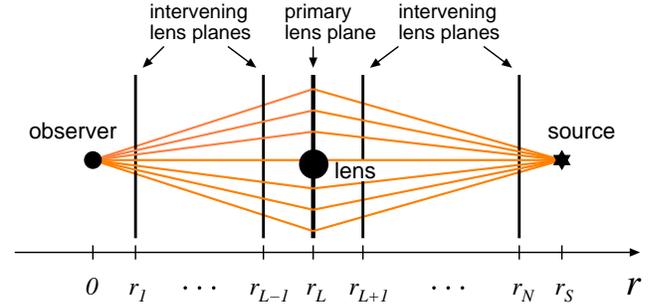}
\caption{
A schematic picture of our ray-tracing simulation.
The horizontal axis is the comoving distance $r$ from the observer.
The vertical thick lines denote the positions of the lens planes
 at $r_j=L_{\rm box} \times (j-1/2)$ with $j=1,2,\cdot \cdot, L,
 \cdot \cdot, N$.
The primary lens plane is located at $j=L$, and the others correspond
 to intervening lens planes.
The source plane is placed at $r_s=L_{\rm box} \times N$.
Here, we set $L_{\rm box}=10h^{-1}$Mpc.
The orange lines are unperturbed
 light rays from the source to
 the observer.
We consider  $1000^2$ light rays that converge at the observer within
 the field-of-view of $38.4^2\,{\rm arcsec}^2$.
}
\label{fig_lens_planes}
\vspace*{0.5cm}
\end{figure}

We carry out ray-tracing 
simulations to calculate the perturbed convergence,
 shear and deflection angle due to intervening structures.
Fig. \ref{fig_lens_planes} shows a schematic picture of our ray-tracing
simulation. The horizontal axis denotes the comoving distance $r$ from the
observer: The thick vertical line is the primary lens plane, and
 the thin vertical lines correspond to the intervening lens planes.
The lens planes are placed at an equal distance interval of $L_{\rm box}$,
 $r_j=L_{\rm box} \times (j-1/2)$ with an integer $j=1,2,\cdot \cdot,
 L, \cdot \cdot, N$.
Here, $j=L$ corresponds to the primary lens plane.
The source plane is placed at $r_S=L_{\rm box} \times N$.
The interval of the lens planes is set to $L_{\rm box}=10h^{-1}\,$Mpc.
We prepare $520$ lens planes up to the source redshift $z_s=4$. 
The orange lines are light-ray paths from the source to the observer,
which are deflected at the primary lens plane.
We use the Born approximation in which the lensing quantities such
 as the convergence, shear, deflection angle 
 are evaluated along unperturbed paths.

\subsection{$N$-body simulations}

We run $N$-body simulations on a cubic box with sides $L_{\rm box}$,
 and then project the
positions of particles to a plane at each redshift bin, 
and obtain the
particle distribution and gravitational potentials on intervening
lens planes by considering randomly oriented observers at each redshift
bin. Note that the effects of large-scale correlation 
between different snapshots are suppressed due to the randomisation of 
orientation. As long as we concern with structures 
whose sizes are much less than the size of the cubic box, such 
effects are negligible. 
We use the numerical $N$-body simulation code Gadget2
 \citep{springel2001, springel2005}.
The initial conditions of our simulations are calculated
using the second-order Lagrangian 
perturbation theory (2LPT) \citep{crocce2006, nishimichi2009}
with an initial linear power spectrum \citep{eisenstein1999}.
We dump the outputs (the particle positions) at the redshifts
that correspond to the positions of lens planes $r_j=L_{\rm box} \times
 (j-1/2)$, as shown in Fig. \ref{fig_lens_planes}.

We run a high and a low resolution simulations to check our numerical 
 convergence at small scales.
In the high (low) resolution simulation, we employ $1024^3(512^3)$
 dark matter particles on a simulation box with 
the sides of $L_{\rm box}=10h^{-1}\,$Mpc
 and the softening length is set to $2\%$ of the mean particle separations
 for the both cases.
The particle mass is $7.1 \times 10^4 (5.7 \times 10^5) h^{-1} M_\odot$
 for high (low) resolution case.
We prepare a single realisation (two realisations) for the high (low)
 resolution simulation.
The simulation data are the same as that used in our previous paper (IT12).
The matter power spectra of our $N$-body simulations agree with the
results of the simulation with higher resolution in which we use
finer simulation parameters for the time step, force calculation,
etc., within $2(6) \%$ for $k < 100 (320) h {\rm Mpc}^{-1}$.

Before closing this section, we comment on the effects of our smaller
 simulation box on our results.
Our simulation box is very small ($L_{\rm box}=10h^{-1}$Mpc on a side)
 and hence
 it does not include the larger fluctuations than the box size which 
 may affect the small-scale clustering and even worse the whole box may
 enters in the non-linear regime.
However, our simulation results can be trusted because of the following reasons:
i) As shown in Fig.2 in our previous paper (IT12), the computed matter power
 spectra for a larger box ($L_{\rm box}=320h^{-1}$Mpc) can be smoothly 
connected to the power spectra for the small one ($L=10h^{-1}$Mpc) 
at $k \sim 30h {\rm Mpc}^{-1}$ for redshifts $z=0-2.2$. This suggests 
that our simulations successfully reproduce the 
clustering of dark matter at smaller scales $< 200 h^{-1} {\rm pc}
 \, (k/30h {\rm Mpc}^{-1})^{-1}$.
ii) To check the non-linearity in the whole simulation box, we evaluate the
 mass variance in the box analytically\footnote{The mass variance within the
 cubic box of size $L$ is given by,
\BE
 \sigma^2_{\rm box}(z;L) = \int\! \frac{d^3 k}{(2 \pi)^3} \left|
 \tilde{W}(\K;L) \right|^2 P_\delta(k,z),
\EE
where we use the top-hat window function of the cubic box as,
 $W(\x;L)=1/L^3$ for $|x|<L/2, |y|< L/2$ and $|z|<L/2$, and $W=0$ otherwise.
$\tilde{W}$ is its Fourier component
 given by $\tilde{W}(\K;L) = {\rm sinc}(k_x L/2) \,{\rm sinc}(k_y L/2)
 \,{\rm sinc}(k_z L/2)$ with ${\rm sinc}(x)=\sin(x)/x$.
For the matter power spectrum $P_\delta(k,z)$, we use the non-linear fitting
 formula given in IT12.}.
The standard deviations are $\sigma_{\rm box}=0.39, 0.30, 0.23$ and $0.15$
 at redshifts $z=0, 0.5, 1$ and $2$.
Note that the most important intervening structures reside at relatively
 higher redshifts $z \gtrsim 0.5$, and hence the variance is smaller than
 unity. This suggests that the simulated particles in the small box are
 in the linear regime.

\subsection{Ray-tracing simulations}
\label{3.2}

We briefly explain our procedure to trace light rays through 
the obtained $N$-body data.
We use a publicly available code called RAYTRIX (Hamana \& Mellier 2001) which
 follows the standard multiple lens plane algorithm\footnote{see
 http://th.nao.ac.jp/MEMBER/hamanatk/RAYTRIX/}.
The distance between an observer and a source is divided into a lot of
 intervals.
As shown in Fig. \ref{fig_lens_planes}, we adopt a
fixed interval of $L_{\rm box}$. 
Particle positions are projected onto two dimensional lens planes
 ($xy$, $yz$ ,$zx$ planes) at every interval $L_{\rm box}$ on 
each light ray path.
 Using the Triangular-Shaped Cloud (TSC) method (Hockney \& Eastwood 1988), 
 we assign the particles onto $N_g^2$ grids in the lens planes, then compute
 the projected density contrast at each plane. We test the convergence
 of our simulations by varying resolution as $N_g^2=2048^2, 4096^2, 8192^2$
 and $16384^2$ which correspond to the grid sizes 
 $r_{\rm grid} \equiv L_{\rm box}/N_g=4.8, 2.4, 1.2$ and
 $0.6 \ h^{-1}$kpc, respectively. 
The two-dimensional gravitational potential is solved via the Poisson
 equation using Fast Fourier Transform. Finally, the 
two dimensional sky maps of 
the convergence, shear, and deflection angle of light rays
are obtained by solving the evolution
equation of the Jacobian matrix along the unperturbed light-ray paths.

We prepare $100$ realisations for each lens system by randomly choosing
 the projected direction and shifting the two dimensional projected
 positions. 
In each realisation, we emit $1000^2$ light-rays in the field of view
 of $38.4 \times 38.4 \ {\rm arcsec}^2$, and the resulting angular
 resolution is $0.04$ arcsec.

\begin{figure}
\vspace*{1.0cm}
\includegraphics[width=85mm]{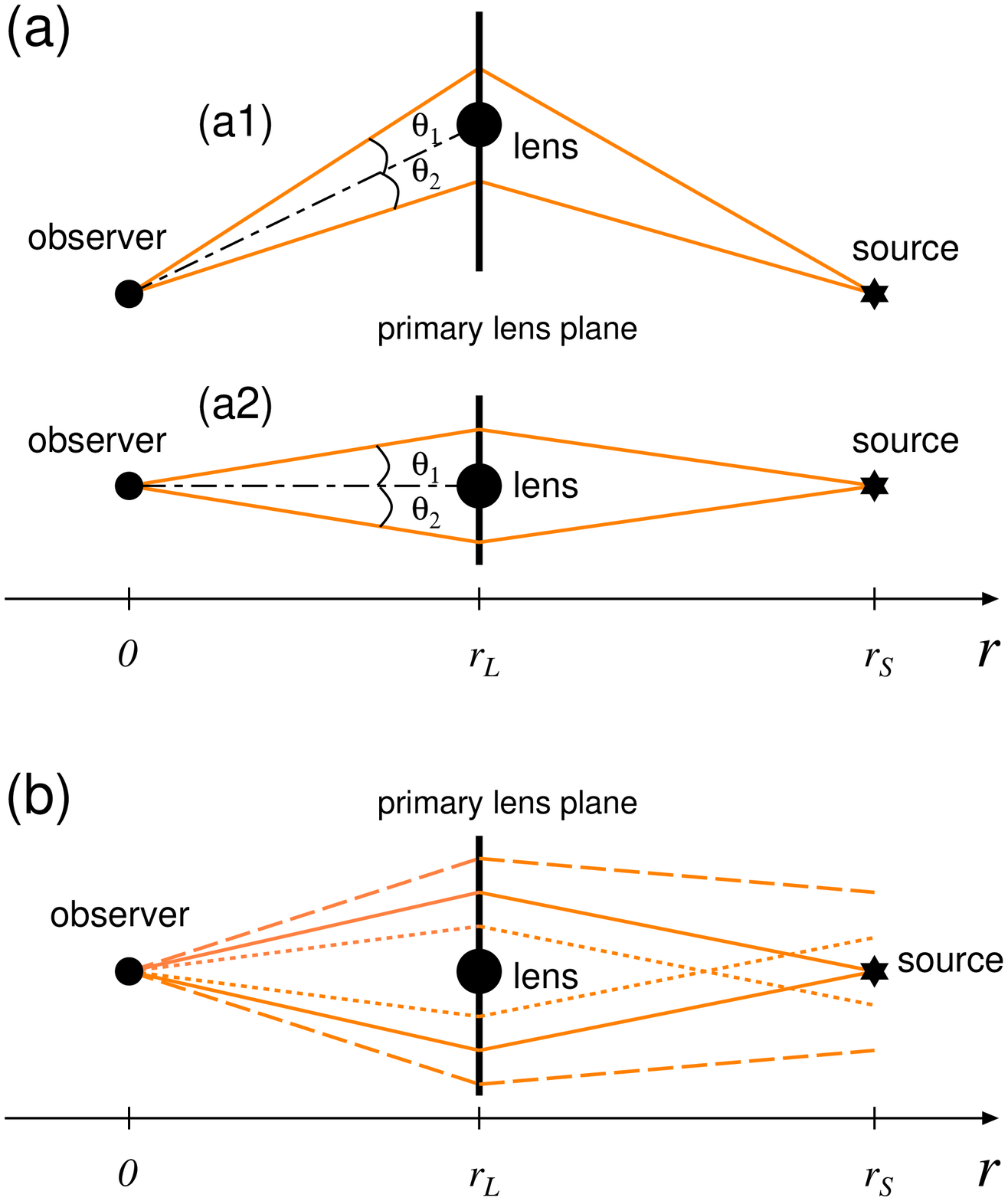}
\caption{
Same as Fig. \ref{fig_lens_planes}, but show two lens systems in panel
 (a) and ``exact'' light-rays in panel (b).
In panel (a), a lens system (which consists of a lens and two images) is put
 on two different positions on the primary lens plane, denoted as
 (a1) and (a2).
The relative image positions $\THE_1$ and $\THE_2$ as measured
 from the lens center for (a1) are the same
 as those for (a2), respectively. 
In panel (b), the orange solid lines are the light-rays that converge
at the position of the observer.
The dotted (dashed) lines are the light-rays that are closer to (far from)
 the lens at the lens plane. }
\label{fig_lens_planes6}
\vspace*{0.5cm}
\end{figure}

Before closing this section, we comment on our simplifications
for implementing fast-calculation. 
First, instead of changing the position of a source, we
change the position of the center of a lens 
(there are $1000^2$ points in single realisation)
as shown in panel (a) of Fig. \ref{fig_lens_planes6}.
In panels (a1) and (a2), a lens system (which consists of a lens and
 double images at angular separations $\THE_{1,2}$ from the lens center)
 is put on different points on the primary lens plane.
As long as the angular separation between lensed images 
is much smaller than the distance from the observer to the source, the 
statistics of the perturbed convergence and shear do not depend on
the position of the lens because of the spatial homogeneity of the 
perturbations. In order to reduce the calculation time, 
we choose all the grid points as the lens centers for each realisation
in calculating the probability distribution function of magnification 
perturbation.

Next, in order to reduce the computation time, we use only photon paths that have a ``diamond shape'' centered at the
lens center as shown in 
orange lines in Fig. \ref{fig_lens_planes}. 
In a real setting, however, light rays passing sufficiently close to the
 lens intersect each other before reaching backward to the source (as shown
 by the dotted lines in Fig. \ref{fig_lens_planes6}),
 whereas the light rays passing far from the lens does not converge
 completely at the source plane (the dashed lines in
 Fig. \ref{fig_lens_planes6}).
The light-ray paths as shown in Fig. \ref{fig_lens_planes} are correct only
for a circular symmetric lens where the source is put at the lens center.
Any deviation from this setting cause errors in calculating statistical 
quantities. 

\begin{table*}
\hspace{-4.5cm}
\begin{minipage}{136mm}
\caption{MIR lens systems}
\setlength{\tabcolsep}{2pt}
\begin{tabular}{llccccccc}
\hline
\hline
  lens system  & image(type) & position(arcsec) & flux ratio(obs.) &
   $\kappa$ & $\gamma_1$ & $\gamma_2$ & $\mu$ & flux ratio(model)  \\ 
\hline
  B1422+231 & A(I) & $(-0.385\pm0.000,0.317\pm0.000)$ & A/B=$0.94\pm0.05$ &
   $0.384$ & $0.251$ & $-0.408$ & $6.669$ & A/B=$0.797$ \\ 
  $~z_L=0.34$ & B(II) & $(0.,0.)$ && $0.469$ & $-0.098$ & $-0.626$ &
   $-8.370$ & \\
  $~z_S=3.62$ & C(I) & $(0.336\pm0.003,-0.750\pm0.003)$ & C/B=$0.57 \pm 0.06$ &
   $0.364$ & $-0.401$ & $-0.099$ & $4.293$ & C/B=$0.513$ \\
  & D(II) & $(-0.948\pm0.004,-0.802\pm0.003)$ && $1.859$ & $-0.705$ & $1.897$ &
   $-0.298$ & \\
  & G & $(-0.742\pm0.003,-0.656\pm0.004)$ &&&&&& \\
\hline
  MG0414+0534 & A1(I) & $(-0.600\pm0.003,-1.942\pm0.003)$ && $0.509$ & $-0.228$ &
    $-0.360$ & $16.593$ & \\ 
  $~z_L=0.96$ & A2(II) & $(-0.732\pm0.003,-1.549\pm0.003)$ & A2/A1=$0.919 \pm 0.021$ & $0.548$ &
    $-0.486$ & $-0.161$ & $-17.233$ & A2/A1=$1.039$ \\
  $~z_S=2.639$ & B(I) & $(0.,0.)$ & B/A1=$0.347 \pm 0.013$ & $0.464$ & $0.145$ &
    $0.288$ & $5.456$ & B/A1=$0.329$ \\
  & C(II) & $(1.342\pm0.003,-1.650\pm0.003)$ && $0.660$ & $-0.409$ & $0.564$ & $-2.704$ & \\
  & G & $(0.472\pm0.003,-1.277\pm0.003)$ &&&&&& \\
  & X & $(0.857\pm0.011,0.180\pm0.009)$ &&&&&& \\
\hline
  H1413+117 & A(II) & $(0.,0.)$ && $0.568$ & $0.437$ &
    $-0.388$ & $-6.430$ & \\ 
  $~z_L=1.88(\star)$ & B(I) & $(-0.744\pm0.003,0.168\pm0.003)$ & B/A=$0.84 \pm 0.07$ & $0.481$ &
    $-0.101$ & $0.288$  & $5.748$ & B/A=$0.894$ \\
  $~z_S=2.55$ & C(I) & $(0.492\pm0.003,0.713\pm0.003)$ & C/A=$0.72 \pm 0.07$ & $0.458$ &
    $-0.347$ & $0.045$ & $5.820$ & C/A=$0.905$ \\
  & D(II) & $(-0.354\pm0.003,1.040\pm0.003)$ & D/A=$0.40 \pm 0.06$ & $0.654$ & $0.454$ &
    $-0.503$ & $-2.948$ & D/A=$0.458$ \\
  & G & $(-0.142\pm0.003,0.561\pm0.003)$ &&&&&& \\
  & X & $(-1.87\pm0.07,4.14\pm0.07)$ &&&&&& \\
\hline
  PG1115+080 & A1(I) & $(-1.328\pm0.003,-2.034\pm0.003)$ && $0.502$ & $-0.188$ & $0.369$ &
    $13.099$ & \\ 
  $~z_L=0.31$ & A2(II) & $(-1.477\pm0.004,-1.576\pm0.003)$ & A2/A1=$0.93 \pm 0.06$ & 
    $0.546$ & $-0.526$ & $0.113$ & $-11.950$  & A2/A1=$0.912$ \\
  $~z_S=1.72$ & B(II) & $(0.341\pm0.003,-1.961\pm0.003)$ && $0.602$ & $-0.101$ & $-0.704$ &
    $-2.87438$ & \\
  & C(I) & $(0.,0.)$ && $0.387$ & $0.309$ & $0.088$ & $3.66826$ & \\
  & G & $(-0.381\pm0.003,-1.344\pm0.003)$ &&&&&& \\
\hline
  Q2237+030 & A(I) & $(0.,0.)$ && $0.394$ & $0.388$ & $0.078$ & 
    $4.733$ & \\ 
  $~z_L=0.04$ & B(I) & $(0.673\pm0.003,1.697\pm0.003)$ & B/A=$0.84 \pm 0.05$ & 
    $0.375$ & $0.086$ & $0.380$ & $4.197$ & B/A=$0.887$ \\
  $~z_S=1.695$ & C(II) & $(-0.635\pm0.003,1.210\pm0.003)$ & C/A=$0.46 \pm 0.02$ & $0.743$ &
    $-0.555$ & $-0.480$ & $-2.117$ & C/A=$0.447$  \\
  & D(II) & $(0.866\pm0.003,0.528\pm0.003)$ & D/A=$0.87 \pm 0.05$ & $0.636$ & $-0.367$ &
    $-0.504$ & $-3.904$ & D/A=$0.825$ \\
  & G & $(0.075\pm0.004,0.939\pm0.003)$ &&&&&& \\
\hline
  RXJ1131-1231($\star \star$) & A(II) & $(-0.588\pm0.003,1.120\pm0.003)$ & A/B=$1.63^{+0.04}_{-0.02}$ &
    $0.464$ & $-0.512$ & $0.274$ & $-20.276$ & A/B=$1.657$ \\ 
  $~z_L=0.295$ & B(I) & $(-0.618\pm0.003,2.307\pm0.003)$ && $0.420$ & $-0.465$ & $-0.196$ &
    $12.239$ & \\
  $~z_S=0.658$ & C(I) & $(0.,0.)$ & C/B=$1.19^{+0.03}_{-0.12}$ & $0.441$ &
    $-0.049$ & $0.470$ & $11.124$ & C/B=$0.909$ \\
  & D(II) & $(2.517\pm0.003,1.998\pm0.003)$ && $0.886$ & $-0.862$ & $0.510$ & $-1.011$
    & \\
  & G & $(1.444\pm0.008,1.706\pm0.006)$ &&&&&& \\
\hline
\label{table1}
\end{tabular}
\\ Note: ($\star$): The lens redshift $z_L$ is obtained from a best-fit model using 
the observed positions of the images and the primary lens, the flux ratios, and
the time-delays between the images assuming
 $H_0=70\,\textrm{km}/\textrm{s}/\textrm{Mpc}$. 
 ($\star \star$): [OIII] line flux ratios.
\end{minipage}
\end{table*}

\section{MIR Quadruple lenses}

In this paper, we use the observed 
MIR (mid-infrared) flux ratios in six lens systems:
 B1422+231, MG0414+0534, H1413+117, PG1115+080, Q2237+030 and RXJ1131-1231.
We use the observed flux of [OIII] line for RXJ1131-1231, and those of 
continuum for the other systems.
 Table \ref{table1} summarises the observed data: 
 in the first column, we list 
the names of lens systems with the source and lens redshifts, in the second
 column, the names of images (A,B,C,$\cdots$) with the types (I,II)
 and lens (G),   
 in the third column, the angular positions of lensed images and the
 centroid of each lensing galaxy in terms of
 right ascension and declination (in units of arcsec), in the forth
 column, the MIR flux ratios.
The data of positions of lensed images and centroids of 
lensing galaxies are taken from 
 the CASTLES webpage\footnote{http://www.cfa.harvard.edu/castles/}
 (note that the signs of right ascension in the Table \ref{table1} are
 opposite to those in the CASTLES webpage).
The six lens systems are listed in descending order of the source redshift.

The MIR bands are suitable to probe intervening structures 
with mass scales of $\gtrsim 10^3 M_{\odot}$ assuming that
the size of a MIR emission region around a quasar 
is $\sim 1\,$pc. This is because the Einstein radii of a point mass 
with a mass $\sim 10^3 M_{\odot}$ are $\sim 1\,$pc at 
cosmological distances. On the other hand, 
the Einstein radius of stars residing at  
a lensing galaxy is typically $\sim 0.001\,$pc. Therefore, microlensing effect
due to stars is negligible provided that the 
filamentary structure of hot dust torus is not prominent
\citep{stalevski12} (see also \cite{sluse13}).
Moreover, the dust extinction is very small in the MIR bands.
Hence, the MIR flux ratios provide us variable information about
structures in the line-of-sight or lensing galaxies (e.g. Chiba et al. 2005).

In the following, we shortly describe observed quantities for each lens.
For detail, see IT12.

The cusp-caustic lens B1422+231 consists of three bright images
A, B, C, and a faint image D. 
Images A, C are minima and B, D are saddles.   
It was first discovered by \citet{patnaik1992} using the 
Very Large Array.
The quasar redshift is $z_S=3.62$ and the lensing galaxy is 
 at $z_L=0.34$ \citep{kundic1997b,tonry1998}.
The MIR image fluxes were measured by \citet{chiba2005} using the Subaru
 telescope.
We use the flux ratios using the bright images A, B, and C (A/B and C/B).
When the source is close to and inside a cusp, then the bright images 
A, B, C can form and the flux ratios satisfy the so called cusp
 caustic relation: $R=(\mu_{\rm A}+\mu_{\rm B}+\mu_{\rm C}) /
 (|\mu_{\rm A}|+|\mu_{\rm B}|+|\mu_{\rm C}|) =0$.
However, the relation does not hold in the optical, MIR, and radio bands. 
The observed flux ratios show an anomaly $R \simeq 0.2$ in the MIR
 \citep{chiba2005} and the radio \citep{mao1998} bands.

The fold-caustic lens MG0414+0534 consists of two bright images A1, A2 and 
two faint images B, C. The images A1, B are minima and A2, C are saddles.
A source quasar at $z_S=2.64$ is lensed by an elliptical galaxy at
 $z_L=0.96$ \citep{hewitt92,lawrence1995, tonry1999}.
A simple lens model, a singular isothermal ellipsoid with
an external shear (SIE-ES) can not fit the image positions as well as 
the flux ratios. \citet{schechter1993} and \citet{Ros2000} 
suggested that another galaxy called ``X'' is necessary to 
account for the relative image positions.
\citet{minezaki2009} measured the MIR flux ratios using the Subaru telescope.
Recently, \citet{macleod2012} have provided more accurate MIR flux ratios
using the Gemini(north) telescope.
The observed flux ratios show an anomaly $R \simeq 0.2$ in the MIR
\citep{minezaki2009, macleod2012} and the radio
\citep{Ros2000,trotter2000} bands.
We use the MIR flux ratios among the bright images A1, A2 and B (A2/A1 and
 B/A1) measured by \citet{minezaki2009} and \citet{macleod2012}.

The clover leaf lens H1413+117 consists of four bright images A, B, C and D.
The images B, C are minima and A, D are saddles. 
The source redshift is $z_S=2.55$ \citep{magain1988} but the lens
 redshift is unknown. 
\citet{goicoechea2010} measured the time delays among the images A-D and
 estimated the lens redshift, $z_L=1.88^{+0.09}_{-0.11}$.
\citet{macleod2009} modeled a galaxy ``X'' as a singular isothermal
sphere (SIS) to improve the fitting
 of the image positions and the flux ratios.
We use the MIR flux ratios among the four images A,B,C and D (B/A, C/A and
 D/A) measured by \citet{macleod2009}.

The fold-caustic lens PG1115+080 consists of two bright images A1, A2 and
 two faint images C, D.
A source quasar at $z_S=1.72$ is lensed by a foreground galaxy 
at $z_L=0.31$ \citep{weymann80,kundic1997b}. 
We use the MIR flux ratios between A1 (minimum) and A2 (saddle)
measured by \citet{chiba2005}.

The clover leaf lens Q2237+030 is the nearest lens
in our samples and it consists of four images A, B, C, and D.
The images A, B are minima and C, D are saddles.
The source is located at $z_S=1.695$ and the lens at $z_L=0.0394$
 \citep{Huchra1985}.
We use the flux ratios among the four images A, B, C, and D (B/A, C/A and
 D/A) measured by \citet{minezaki2009}.

The cusp-caustic lens RXJ1131-1231 consists of three bright images A, B,
C, and a faint image D. The quasar at $z_S=1.72$ lensed by a 
foreground galaxy at
 $z_L=0.31$\citep{sluse2003}. 
We use the [OIII] flux ratios of images A (saddle), B (minimum) 
and C (minimum) (A/B and C/B) measured by \citet{sugai2007}.

\section{Method}

\begin{figure*}
\vspace*{1.0cm}
\includegraphics[width=130mm]{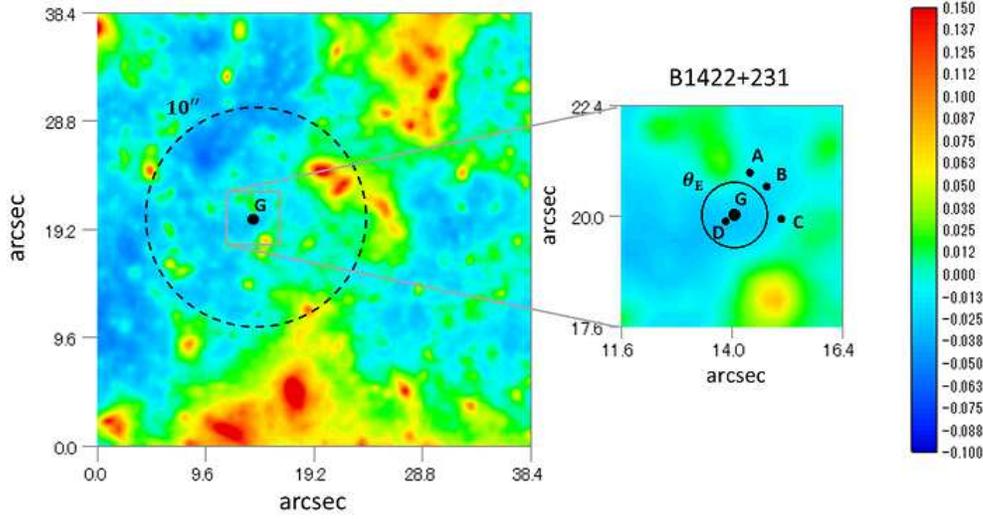}
\caption{
Contour map of convergence perturbation $\delta \kappa$
due to the line-of-sight structures for B1422+231.
The contours in the left panel show 
the total field of view of $38.4 \times 38.4 {\rm sec}^2$,
and those in the right panel show
the enlarged view around a lens. 
The point G is the centroid of the lensing galaxy and 
A, B, C, and D correspond to the image positions.
In the left panel, a dashed circle centered
at G has a radius of $10$ arcsec.  
Similarly, in the right panel, the circle has an effective 
Einstein radius of the lens. The map is obtained 
from a simulation of $1024^3$ $N$-body particles with
a grid radius $r_{\rm grid}=1.2h^{-1}$kpc.
}
\label{fig_kappa_map_B1422}
\vspace*{0.5cm}
\end{figure*}

This section describes our procedure for evaluating the weak lensing
effects caused by the line-of-sight structures on 
the flux ratios of lensed images.
First, we make a primary lens model to fit the observed MIR image 
positions, and then we evaluate the unperturbed
 convergence $\kappa_{\rm i}$ and shear $\gamma_{\rm i}$ for each image.
Next, we obtain the perturbed convergence $\delta \kappa_{\rm i}$ and shear
 $\delta \gamma_{\rm i}$ due to the 
line-of-sight structures using ray-tracing
 simulations, and then we add the perturbed quantities $\delta \kappa_{\rm i},
 \delta \gamma_{\rm i}$ to the original ones $\kappa_{\rm i}, \gamma_{\rm i}$.
Finally, we evaluate the magnification contrast $\delta^\mu_{\rm i}=
 \delta \mu_{\rm i}/\mu_{\rm i}$ and the magnification perturbation 
$\eta$ defined in equation (\ref{eta_def}).

\subsection{Primary lens model}
\label{5.1}

\begin{table*}
\caption{Best-fit model parameters for 6 MIR lens systems (see IT12 for detail)}
\begin{tabular}{lccccccc}
\hline
  lens system  & $b^\prime$ (arcsec) & $b_\textrm{X}$ (arcsec) & $\THE_0$ (arcsec) & $e$ &
   $\theta_{\rm e}$ (deg) & $\gamma$ & $\theta_\gamma$ (deg) \\ 
\hline
 B1422+231 & $0.755$ & & $(-0.741,-0.658)$ & $0.309$ & $-56.6$ & $0.166$ &
 $-52.3$ \\ 
 MG0414+0534 & $1.08$ &0.185& $(0.472,-1.277)$ & $0.232$ & $-82.1$ & $0.102$ &
 $53.8$ \\ 
 H1413+117 & $0.561$ &0.583& $(-0.172,0.561)$ & $0.204$ & $-14.5$ & $0.062$ &
 $55.7$ \\ 
 PG1115+080 & $1.14$ & & $(-0.361,-1.342)$ & $0.156$ & $-83.0$ & $0.110$ &
  $51.8$ \\ 
 Q2237+030 & $0.854$ & & $(0.075,0.939)$ & $0.371$ & $64.9$ & $0.015$ & 
  $-46.8$  \\ 
 RXJ1131-1231 & $1.83$ & & $(2.039,0.568)$ & $0.145$ & $-57.8$ & $0.120$ &
  $-81.8$  \\ 
\hline
\label{table2}
\end{tabular}
\medskip
\end{table*}

First, for each lens system, we fit the observed positions of 
lensed images and the centroid of a lensing galaxy by changing the 
lens parameters of a smooth mass lens model.
Note that we do not include the measured MIR fluxes in the fitting.
For detailed discussion, see IT12.
We use a singular isothermal ellipsoid (SIE) with an external shear (ES)
 \citep[e.g.,][]{skw2006} as a canonical lens model.
The SIE model is characterized by the effective 
Einstein radius $b^\prime$, position of the lens center $\THE_0$, and 
ellipticity $e$ with the position angle $\theta_{\rm e}$.
The external shear is expressed in terms of the amplitude $\gamma$ and the
 direction $\phi_\gamma$.
In terms of these parameters, the lensing potential of 
a SIE plus ES (SIE-ES) can be written as
\BE
 \phi(\THE) = \tilde{\THE} \cdot \ALP(\tilde{\THE}) - \frac{\theta^2}{2}
 \gamma \cos \left[ 2 \left( \phi_\theta - \phi_\gamma \right) \right],
\EE
where $\THE=(\theta_1,\theta_2)=(\theta \cos \phi_\theta, \theta
 \sin \phi_\theta)$ is the angular coordinate from the lens center
 $\THE_0$, and $\tilde{\THE}$ is the coordinate rotated by an angle
 $\theta_{\rm e}$: $\tilde{\theta}_1 = \theta_1 \cos \theta_{\rm e} +
 \theta_2 \sin \theta_{\rm e}$, $\tilde{\theta}_2 = - \theta_1
 \sin \theta_{\rm e} + \theta_2 \cos \theta_{\rm e}$.
The deflection angle $\ALP(\THE)$ of the SIE is given by,
\BE
 \alpha_{1,2}(\tilde{\THE})=\frac{b^\prime q}{\sqrt{1-q^2}} \tan^{-1}
 \left[ \frac{ \sqrt{1-q^2}}{\sqrt{q^2 \tilde{\theta}_1^2 +
 \tilde{\theta}_2^2}} \tilde{\theta}_{1,2} \right].
\EE
where $q=1-e$.
We use the public code
 GRAVLENS\footnote{http://redfive.rutgers.edu/\~{}keeton/gravlens/} by Keeton
 to find the best-fit model parameters.
Note again that we use only the positions of lensed 
images and the centroid of the primary lens to fit the model parameters.
The best-fit parameters for all the systems are summarised
 in Table \ref{table2}.
The resulting convergence $\kappa_{\rm i}$, shear $\gamma_{\rm i}$,
 magnification $\mu_{\rm i}$ for each image based on the best fitting
 model are shown in Table \ref{table1} (from the 5th to 8th columns).
Similarly, the theoretically predicted flux ratios are shown in
 the last column in Table \ref{table1}.
For MG0414 and H1413, object ``X'', modeled by a singular isothermal
 sphere, is added at the positions $(0.857,0.180)$ and $(-1.87,4.14)$
 arcsec, respectively. The effective Einstein radii $b_{\textrm{X}}$ 
are shown in Table \ref{table1}. For detail, see IT12.

\subsection{Weak lensing by line-of-sight structures}
\label{5.2}

As discussed in section \ref{sec3}, 
 we prepare $100$ lensing convergence and shear maps for each
 lens system using our ray-tracing simulations.
The field of view is $38.4 \times 38.4$ ${\rm sec}^2$ with
 $1000^2$ grids, and thus the resulting angular resolution is $0.038$
 arcsec. 
In Fig. \ref{fig_kappa_map_B1422}, we plot an example of the 
convergence perturbation due to the line-of-sight structures for B1422+231.
The left panel and the right panel indicate the total field of view 
of $38.4 \times 38.4 {\rm sec}^2$ and the enlarged view
 around the primary lens, respectively.
The centroid of the lensing galaxy is 
denoted as G and the image positions are A, B, C, and D.
In the maps, red spots correspond to overdense regions, while 
blue spots to underdense regions (or voids).
We can see some massive clumps or filamentary
structures in the neighborhood of the primary lens.
We obtain the perturbed convergence $\delta \kappa_{\rm i}$ and shear
 $\delta \gamma_{{\rm i} 1,2}$ on grid points at image positions.

In what follows, we describe our procedure for evaluating the probability
 distribution of magnification perturbation. \\

1) We put the lens center at an arbitrary grid point in a simulated 
map and also put the images of a point source on the 
observed image positions.
In this procedure, we do not allow to place a halo 
that is more massive than the primary lens in the line-of-sight 
since it should be ``primary''.
As a result, we impose the following two conditions in our analysis:
(i) Convergence perturbations in a $10^{\prime \prime}$ circle around
 the primary lens galaxy is less than $0.5$. The circle is plotted as dashed
 curve in Fig.  \ref{fig_kappa_map_B1422}.
(ii) Convergence perturbations $\delta \kappa$ and shear perturbations 
$\delta \gamma$ should be smaller than those of the primary lens, i.e.
 $|\delta \bar{\kappa}| < |\bar{\kappa}|$ and $|\delta \gamma| <
 |\bar{\gamma}|$. Here, $\delta \bar{\kappa}$ and $\delta \bar{\gamma}$
 are the mean perturbed convergence and shear inside a circle with the
 Einstein radius at the lens center, and $\bar{\kappa}$ and $\bar{\gamma}$
 are the mean convergence and shear among all the images in the lens system.

In our analysis, we include only samples that satisfy
the above two conditions.
Note that almost all the light rays satisfy the above conditions (i)
 and (ii).
The convergence and shear perturbations are usually much smaller than $1$.
For instance, only $3\%(0.2\%)$ rays have $\delta \kappa >
 0.1(0.3)$ in B1422+231.
Since the source redshift of B1422+231 is the highest in the six lens
 systems, the convergence perturbations $\delta \kappa$ in the other
 systems are naturally smaller than that in B1422+231.   \\

2) We obtain the convergence perturbations  $\delta \kappa_{\rm i}$
 and shear perturbations $\delta \gamma_{\rm i}$ for the macro-lensed  
images from our ray-tracing simulation.
However, contribution from 
the external shear and the constant convergence is
 already taken into account in the primary lens.
Therefore, we should subtract the mean shear and convergence
 perturbations around the lens in order to avoid the double-counting.
We calculate the mean convergence and shear perturbations,
 $\delta \bar{\kappa}$ and $\delta \bar{\gamma}_{1,2}$, inside a circle
 with the effective Einstein radius around the center of the lens. 
Then, we subtract them from the original ones: $\delta \kappa_{\rm i}
 \rightarrow
 \delta \kappa_{\rm i}^\prime = \delta \kappa_{\rm i} - \delta \bar{\kappa}$
 and $\delta \gamma_{{\rm i} 1,2} \rightarrow
 \delta \gamma_{{\rm i} 1,2}^\prime = \delta \gamma_{{\rm i} 1,2}  -
 \delta \bar{\gamma}_{1,2}$.

Similarly, we also subtract the deflection angle arising from the mean
 external field. 
The convergence and the shear correspond to the second derivatives of
 the lensing potential while the deflection angle corresponds to 
the first derivative, and hence the 
constant shear corresponds to the deflection angle as a
linear function of the angular coordinates ($\theta_1, \theta_2$).
First, we fit the deflection angle $\ALP(\theta_1,\theta_2)$ as a linear
 function of the angle $(\theta_1,\theta_2)$ inside a circle with the
 Einstein radius.
Since we assume the Born approximation, we evaluate the deflection
 angles along the unperturbed light rays. 
Then, we have $\ALP(\theta_1,\theta_2) \simeq \ALP_0 + \ALP_{11}\theta_1
 + \ALP_{12}\theta_2$, where $\ALP_{0,11,12}$ are constants obtained
 from the fitting.
Then, we subtract the $\ALP_{0,11,12}$ terms from the original deflection
 angle: $\ALP(\theta_1,\theta_2) \rightarrow \ALP^\prime(\theta_1,\theta_2)
 = \ALP(\theta_1,\theta_2) - \ALP_0 - \ALP_{11}\theta_1
 - \ALP_{12}\theta_2$.

The angular positions of lensed images are shifted due to the lensing by
the line-of-sight structures, but we expect that 
the shifts fall within observational errors 
if the image positions are well fit by the primary lens model.
According to CASTLES database,
 the lensed image positions are typically 
measured within errors of $\sim 0.003$ arcsec. 
Then, the errors in the relative positions of any image pairs 
 are $0.0042 (=\sqrt{2} \times 0.003)$ arcsec.
We only use samples in which the maximum shift of the relative angular
 positions among all the lensed images are less than
 $\varepsilon=0.0042 (=\sqrt{2} \times 0.003)$ arcsec in our analysis.
If the shifts of positions due to perturbers in the line-of-sight
are much larger than the observational error $\varepsilon$, we cannot reparametrise
the model parameters to obtain a good fit. Even if one can fit
one of the perturbed positions by reparametrisation, it is impossible
to fit all the positions of lensed images provided that the surface 
density of the lens system has a nearly circular symmetry (see appendix
A). Only allowed shifts are corresponding to
the scale transformation that gives the mass-sheet degeneracy. However, 
we do not exclude the cases in which the order of shifts of images 
are $O(\varepsilon)$ as reparametrisation can improve the shifts by
order $O(\varepsilon)$ in general.

Here, we adopt a circle with the effective Einstein radius around
 the lens as the background (or environment) of the lens. 
However, in real lens systems, the critical curves are elliptical and
 there is an ambiguity how to define the background.
We will discuss these systematic uncertainties in Sec.\ref{systematics}.
 \\

3) For all the images, we add the perturbed quantities as,
 $\kappa_{\rm i} \rightarrow \kappa_{\rm i} +
 \delta \kappa_{\rm i}^\prime$ and $\gamma_{{\rm i} 1,2} \rightarrow
 \gamma_{{\rm i} 1,2} + \delta \gamma_{{\rm i} 1,2}^\prime$, and then 
 calculate the magnification contrast $\delta \mu_{\rm i}/\mu_{\rm i}$
 from equation (\ref{relative_mu}).
 We finally obtain the magnification perturbation $\eta$ from equation (\ref{eta_def}). \\

4) In the previous procedures from 1) to 3), we choose an arbitrary grid
 point as the lens center.
 In order to obtain the probability distribution of $\eta$, we choose all
 the grid points (there are the $1000^2$ points on a single map) as the
 lens centers.
Note again that, to include the lens system in our analysis, we impose a
condition that no massive line-of-sight halo in the
 neighborhood of the lens is allowed.
We do not use the positions that are too close to the edges of the map
 (less than $10^{\prime \prime}$ from the edges) as the lens center, 
in order to calculate the
maximum $\delta \kappa$ within a circle with a radius of 
$10^{\prime \prime}$ centred at the lens center 
(see the condition (i) in the procedure 1)).
To increase the number of realisations, we also use maps rotated
 by $90,180$ and $270$ degrees. Hence, we have 400 maps effectively.

\subsection{Distribution of $\delta \bar{\kappa}$ and $\delta \bar{\gamma}$}

\begin{figure*}
\vspace*{1.0cm}
\includegraphics[width=170mm]{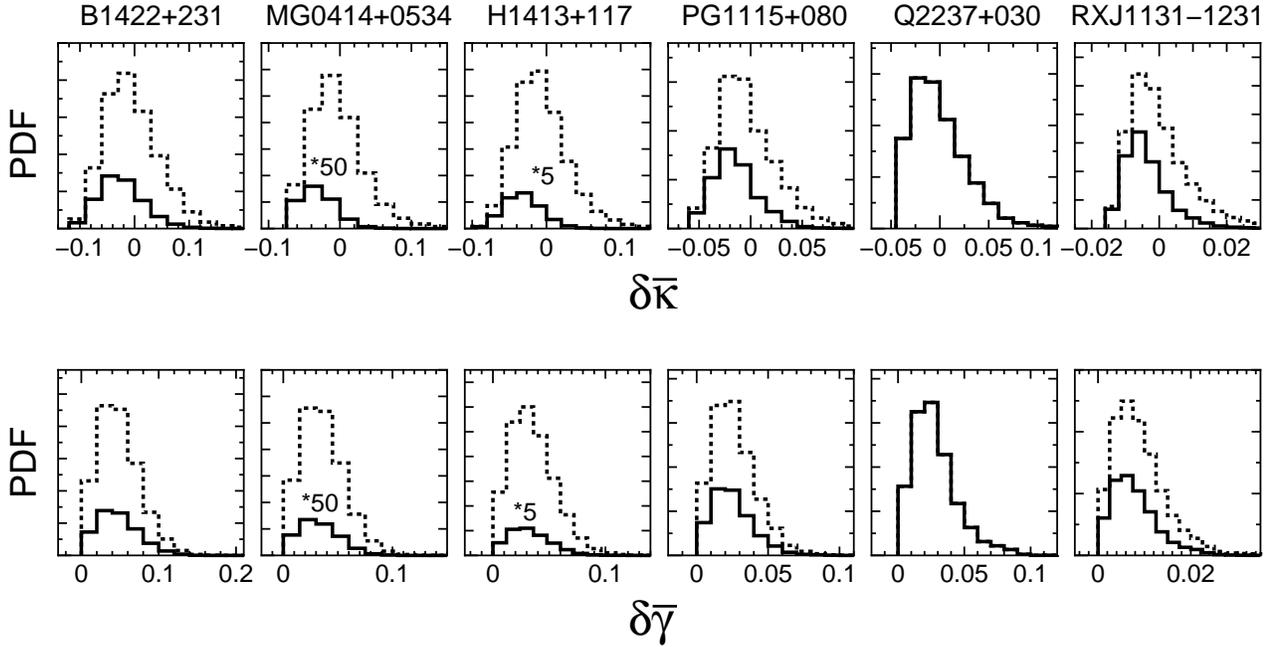}
\caption{
The PDFs of mean perturbed convergence $\delta \bar{\kappa}$ (red)
 and shear $\delta \bar{\gamma}$ (blue) for the six lens systems.
The mean quantities ($\delta \bar{\kappa}$ and $\delta \bar{\gamma}$) are
 smoothed with a circular top-hat filter with the Einstein radius.
The solid (dotted) lines are the PDFs with (without) the condition that 
shifts in the relative positions of pairs of lensed images 
 are less than $0.0042$ arcsec.
Note that the convergence $\delta \bar{\kappa}$ can be positive or negative
 (corresponding to overdense or underdense regions),
 while the shear $\delta \bar{\gamma}$ is positive definite. 
For MG0414+0534 and H1413+117, the amplitudes of these quantities are
 multiplied by factors $50$ and $5$ (full curves).
}
\label{fig_PDF_kappam}
\vspace*{0.5cm}
\end{figure*}

As shown in Figure \ref{fig_PDF_kappam}, the convergence perturbation
$\delta \bar{\kappa}$ is negatively skewed if the shifts in 
the relative positions are constrained to satisfy $\delta \theta
<0.0042$ arcsec. This mean that most of the constrained light-rays pass through 
underdense regions. The variances of PDFs of constrained convergence and shear
perturbations are smaller than those of unconstrained ones, implying
that effects of neighbouring massive structures are suppressed. 
For MG0414+0534 and H1413+117, only a few light rays satisfy the constraint
$\delta \theta <0.0042$ arcsec. 
Since these two lens systems at high redshifts $z_L>0.9$ 
have a large effective Einstein radius
in comoving scale (see Tables \ref{table1} and \ref{table2}), and
large separation angles between lensed images, the 
projected density fluctuations induce 
larger shifts in lensed image positions (see also sec.\ref{analytical}).
As a result, only a few light rays satisfy the constraint.
It can be interpreted that object Xs in these systems may 
reside in intergalactic spaces rather than in the primary lenses. 
In fact, \citet{peng2004}
argues that much of the optical flux of object X in MG0414+0534 
is actually coming from lensed images of the quasar host galaxy rather
than object X itself. Our findings support this interpretation. 
For H1413+117, the perturber(s) may be a void(s) rather than a halo(s)
in the intergalactic space since
observed fluxes of images (B and C) with a positive parity are
demagnified (Table 1).

\section{Analytical formulation}
\label{analytical}

In this section, we present analytical formulation 
for estimating the weak-lensing effects 
by structures in the line-of-sight (see also IT12). The second moment
of the magnification perturbation $\eta$ can be calculated as follows.

The perturbed convergence $\delta \kappa$ and shear $\delta
 \gamma_{1,2}$ caused by density fluctuations
along the line-of-sight are functions of a separation angle $\THE$ between
a pair of lensed images. The two-point correlation function
of these perturbed quantities takes similar forms
 as those for evaluating the weak lensing effect by large-scale structures
 (e.g. Bartelmann \& Schneider 2001). However, 
the unperturbed light-ray paths form a ``diamond shape'' connecting 
 the observer and the source as shown in the orange lines in
 Fig. \ref{fig_lens_planes} because of strong lensing. Contribution 
from large-scale modes should be suppressed as it gives excess astrometric
shifts between pairs of images. They should be constrained from errors 
in the relative positions of lensed images. 

In a cosmological model with the Hubble constant $H_0$ and present matter 
density parameter $\Omega_{m,0}$, the constrained two-point correlation of $\delta \kappa$ as a
 function of a separation angle $\THE$ is 
\BEA
\xi_{\kappa \kappa}(\THE) &\equiv&
\langle \delta \kappa (0) \delta \kappa (\THE) \rangle  \nonumber \\
&=&\frac{9 H_0^4 \Omega_{m,0}^2}{4 c^4}
\int_0^{r_S} dr  r^2 \biggl(\frac{r-r_S}{r_S} \biggr)^2 [1+z(r)]^2
\nonumber \\
&&\times \int_{k_{\rm {lens}}}^{\infty}\frac{dk}{2 \pi} k
 W^2(k)  P_{\delta}(k,r)
 J_0(g(r) k\theta),  \nonumber \\
\label{eq:ck}
\EEA
where
\BE
g(r)= \left\{ 
\begin{array}{ll}
r, & \mbox{$r<r_L$} \\
{r_L(r_S-r)}/{(r_S-r_L)}, & \mbox{$r\ge r_L$}
\end{array}
\right.
\label{eq:g}
\EE
and $P_\delta(k,r)$ is the power spectrum of dark matter 
density fluctuations as a function of the wavenumber $k$ and the
 comoving distance $r$. $r_S$ is the comoving distance to the source
and $r_L$ to the lens from the observer and $z(r)$ is the redshift of a 
point at a comoving distance $r$. 
To calculate $P_\delta$, we use the
fitting function obtained from a high resolution
 cosmological simulation (Appendix of IT12, see also
 \cite{smith2003,takahashi2012}).
The fitting function can be used up to a wavenumber 
 $k = 300\,h {\rm Mpc}^{-1}$ within $\sim 20\%$ accuracy.
$J_0$ is the zero-th order Bessel function and $g(r) \theta$ is
 the tangential separation between two unperturbed light-rays at a
 comoving distance $r$ from the observer.
Strictly speaking, this expression is not exact for
sources that are shifted from the lens center. However,
in what follows, we assume that the angular separation between
the source and the lens center is sufficiently smaller than
the effective Einstein radius of the primary lens so that 
the equation holds with good accuracy. The infrared cut-off
scale in the wavenumber  $k_{\rm {lens}}$ is given by the 
mean separation angle $b$ between a lensed image and
the lens center. For lens systems in which the primary lens
dominates the lensing potential, it approximately equals to the
effective Einstein radius $b_{\rm E}$.  We set the corresponding
wavelength as $\lambda_{\rm lens}=4 b$ or equivalently, $k_{\rm
lens}=\pi/2b$. Any modes whose fluctuation scales are larger than
$\lambda_{\rm lens}$ contribute to the smooth component of the 
primary lens, namely, the constant convergence, shear and possible 
higher order ($m=3,4$) components. Therefore, we only consider 
fluctuation modes whose wavenumbers satisfy $k>k_{\rm lens}$. 
Otherwise, double-counting of the constant convergence and shear
leads to a systematically large perturbation.

The window function $W(k)$ has two cut-off 
scales in wavenumbers $k_{\rm cut}$ and $k_{\rm grid}$.
The cut-off scale $k_{\rm cut}$ is determined
by the condition that the perturbation $\varepsilon$ 
of angular separation between an arbitrary 
pair of lensed images should not exceed the observational error
$\varepsilon_{\rm{obs}}$ for the maximum separation angle between lensed images.
We consider the following two types of filtering. 
The first type is called the ``sharp k-space (SK)'' filter (see also IT12)
in which the filter function satisfies
\BE
  W_{\rm{SK}}(k;k_{\rm cut}) = \left\{ 
\begin{array}{ll}
0, & \mbox{$k<k_{\rm cut}$} \\
1, & \mbox{$k \ge k_{\rm cut}$}.
\end{array}
\right.
\EE
Although the  ``sharp k-space'' filtering is simple and easy for
implementation, contributions from modes with $k_{\rm lens}<k<k_{\rm
cut}$ may not be negligible. Therefore, we introduce another 
type of filtering so called the ``constant shift (CS)'' filter,
\BE
  W_{\rm{CS}}(k;k_{\rm cut}) = \left\{ 
\begin{array}{ll}
W_{\rm{int}}(k), & \mbox{$k<k_{\rm cut}$} \\
1, & \mbox{$k \ge k_{\rm cut}$},
\end{array}
\right.
\EE
\begin{figure}
\vspace*{1.0cm}
\includegraphics[width=90mm]{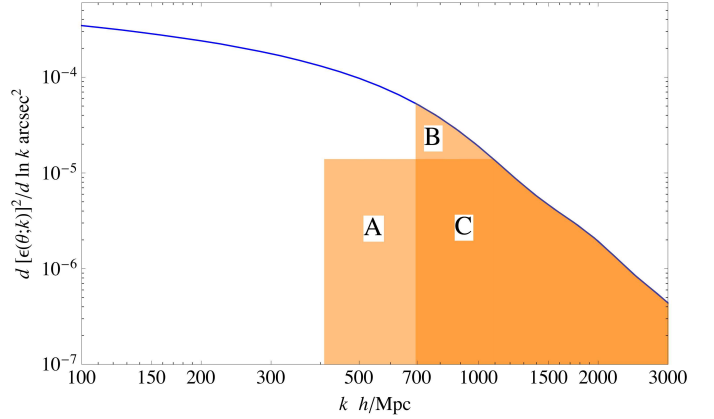}
\caption{Window functions $W(k)$ for B1422+231. 
The upper boundary of the regions B and
C corresponds to the values convolved with the 
sharp k-space (SK) filter with $k_{\textrm{cut}}=694
 h/\textrm{Mpc}$ and those of the regions A and B, to the values
convolved with the 
constant-shift (CS) cut with $k_{\textrm{cut}}=1089  h/\textrm{Mpc}$. Here we assume 
$k_{\textrm{lens}}=408 h/\textrm{Mpc}$. }
\label{CSSKwindow}
\vspace*{0.5cm}
\end{figure}
in which the corresponding contribution to the angular shifts between
a pair of images with the maximum separation angle $\theta_{\max}$ are constant
in logarithmic interval in $k$ for $k<k_{\rm cut}$ 
(see Fig. \ref{CSSKwindow}). 
In this model, contribution from modes with $k_{\rm lens}<k<k_{\rm
cut}$ to the angular shift $\varepsilon$ does not depend on the wavenumber 
$k$ (see also Appendix B for verification).  $W_{\rm{int}}$ is explicitly given by
\BE
W^2_{\rm{int}}(k;k_{\rm cut})\equiv \frac{\del \varepsilon^2(W=1)}{\del \ln{k}}
\bigg /
\frac{\del\varepsilon^2}{\del \ln{k}}\bigg |_{k=k_{\rm cut}},
\EE
where
\BE
\varepsilon^2=2 \langle \delta\theta^2 (0) \rangle - 
2 \langle \delta\theta (0)\delta \theta(\theta_{\rm{max}}) \rangle, 
\EE
and
\BEA
\langle \delta \theta (0) \delta \theta (\theta) \rangle 
&=&
\frac{9 H_0^4 \Omega_{m,0}^2}{ c^4}
\int_0^{r_S} dr  \biggl(\frac{r-r_S}{r_S} \biggr)^2 [1+z(r)]^2
\nonumber
\\
&\times& \int_{k_{\rm {lens}}}^{\infty}\frac{dk}{2 \pi k} W^2(k) P_{\delta}(k;r) J_0(g(r) k\theta).
\nonumber
\\
\label{eq:shiftsq}
\EEA
\begin{figure*}
\vspace*{1.0cm}
\includegraphics[width=180mm]{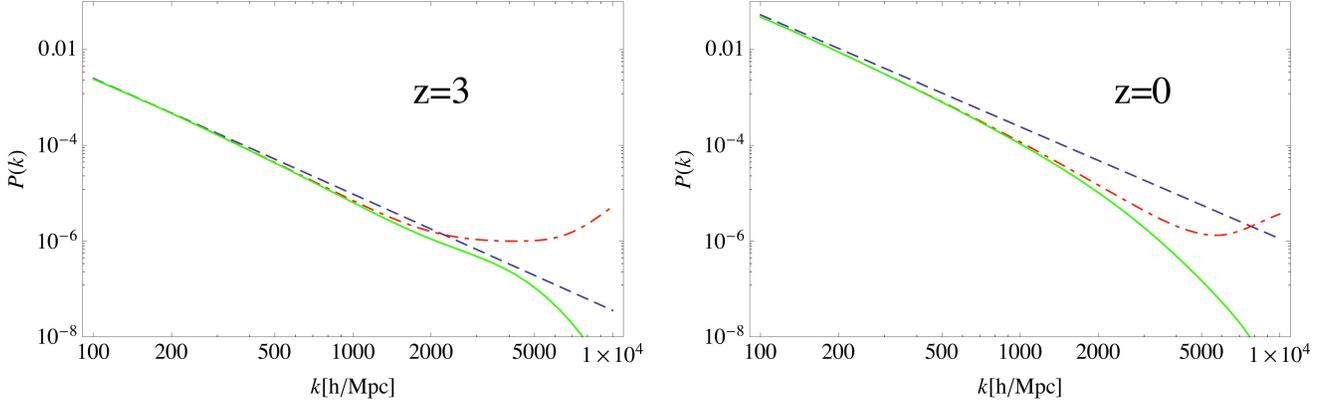}
\caption{Power spectra $P(k)$ for $z=3$ (left) and $z=0$ (right). The plots
of curves represent the fitting function in IT12 (blue dashed),
an $N$-body simulation with $N_p^3=1024^3$ divided by the Cloud In Cell
 (CIC) window function, and the fitting
 function with the TSC smoothing with a grid size
 $r_{\textrm{grid}}=0.6\, h/\textrm{Mpc}$ (green full). 
An increase at small scales for the $N$-body simulation is due to
the Poisson (shot) noise ($1/n=9.3 \times 10^{-7} (N_p/1024)^{-3} h^{-3}
 {\rm Mpc^3}$) and the division by the CIC window function. }

\label{powerTSC}
\vspace*{0.5cm}
\end{figure*}

The angular resolution of ray-tracing simulation 
is given by the grid size $r_{\rm grid}$ of the two dimensional 
gravitational potential. Fluctuations with angular sizes smaller than
$r_{\rm grid}$ are strongly suppressed. For the TSC smoothing scheme, the 
input power spectrum is suppressed by 
a factor $\sim W^2_{\rm TSC iso}(k;k_{\rm grid})=\exp ( -\pi^2 k^2/ k_{\rm grid}^2 )$, where 
$k_{\rm grid}=2 \pi/r_{\rm grid}$ (see Appendix for derivation). As
shown in Fig. \ref{powerTSC}, the 
cut-off scale in comoving coordinates becomes small as the redshift of
fluctuations decreases. This can be explained as follows.
The comoving size of an arc that subtends a given angle become large
as the distance to the arc increases. Therefore, if the angular 
size of the grids is fixed, the corresponding comoving size of 
grids becomes smaller as the distances to the grids decrease.
Incorporating the finite grid-size effect 
for the TSC smoothing, the two types of 
filter function can be written as 
\BE
  W^2(k;k_{\rm cut},k_{\rm grid})= \left\{ 
\begin{array}{ll}
W^2_{\rm{SK}}(k;k_{\rm cut}) W^2_{\rm TSC iso}(k;k_{\rm grid})\, & \mbox{(SK)} \\
W^2_{\rm{CS}}(k;k_{\rm cut}) W^2_{\rm TSC iso}(k;k_{\rm grid})\, & \mbox{(CS)}.
\end{array}
\right.
\EE

The two-point correlation functions for the other perturbed quantities are the 
same as in equation (\ref{eq:ck}) except for the $J_0$ term in the integrand:
\BEA
 \langle \delta \gamma_1(0) \delta \gamma_1(\THE) \rangle &:& J_0 \rightarrow
  \frac{1}{2} \left[ J_0 + J_4 \cos(4 \phi_\theta) \right], \nonumber \\
 \langle \delta \gamma_2(0) \delta \gamma_2(\THE) \rangle &:& J_0 \rightarrow
  \frac{1}{2} \left[ J_0 - J_4 \cos(4 \phi_\theta) \right], \nonumber \\
 \langle \delta \kappa(0) \delta \gamma_1(\THE) \rangle &:& J_0 \rightarrow
  - J_2 \cos(2 \phi_\theta), \nonumber \\
 \langle \delta \kappa(0) \delta \gamma_2(\THE) \rangle &:& J_0 \rightarrow
  - J_2 \sin(2 \phi_\theta), \nonumber \\
 \langle \delta \gamma_1(0) \delta \gamma_2(\THE) \rangle &:& J_0 \rightarrow
  \frac{1}{2} J_4 \sin(4 \phi_\theta), \label{other_corr}
\EEA
where the Bessel functions $J_{0,2,4}$ are functions of $g(r)k\theta$ and
 $\THE=(\theta \cos \phi_\theta, \theta \sin \phi_\theta)$.
When the separation angle $\theta$ between a pair of lensed images 
is sufficiently smaller
than $\theta_E$, the terms which are proportional to 
$J_2$ or $J_4$ are sufficiently smaller than $J_0$. In this case, 
we can drop theses terms and hence from
 equations (\ref{eq:ck}) and (\ref{other_corr}), we have $\langle \delta \gamma_1
 \delta \gamma_1 \rangle$ $=\langle \delta \gamma_2 \delta \gamma_2 \rangle$
 $=\xi_{\kappa \kappa}/2$ and $\langle \delta \kappa \delta \gamma_1
 \rangle =\langle \delta \kappa \delta \gamma_2
 \rangle = \langle \delta \gamma_1 \delta \gamma_2
 \rangle =0$. 

 From equations (\ref{eta_approx}) and (\ref{eta_def}), we can 
obtain the second moment of $\eta$.  
As an example, let us consider three images with two minima A and C and 
 one saddle B with $\kappa_B<1$. 
Choosing coordinates where the separation angle is perpendicular to $+$
mode (i.e., $\theta \sin \phi_\theta=0$), we have $\langle \delta \kappa \delta \gamma_2
 \rangle = \langle \delta \gamma_1 \delta \gamma_2
 \rangle =0$. Then, for $|\delta_i^\mu|\ll1$, the second moment $\langle \eta^2 \rangle$ can
 be written as
\BEA
\langle \eta^2
\rangle&=&\frac{1}{4}\biggl[(I_A+I_B)-2I_{AB}(\theta_{AB})+(I_B+I_C)
\nonumber
\\
&-& \!\!\!\! 2J_{BC}(\theta_{BC}) \biggr],
\label{eq:estimator-anal}
\EEA
where
\BE
I_i\equiv \mu_i^2(4(1-\kappa_i)^2+2 \gamma_{1i}^2 + 2\gamma_{2i}^2)
\xi_{\kappa}(0) ,
\label{eq:Ii}
\EE
and
\BEA
I_{ij}(\theta)\!\!&\equiv&\!\!
4 \mu_i \mu_j \biggl[(1-\kappa_i)(1-\kappa_j)\xi_\kappa(\theta)
\nonumber
\\
&+&
\gamma_{1i}\gamma_{1j}\langle \delta \gamma_1 (0) \delta \gamma_1
(\theta) \rangle
+\gamma_{2i}\gamma_{2j}\langle 
\delta \gamma_2 (0) \delta \gamma_2(\theta) \rangle
\nonumber
\\
&+&
(1-\kappa_i)\gamma_{1j} \langle \delta \kappa_i(0) \delta \gamma_{1j}
(\theta) \rangle
\nonumber
\\
&+&
(1-\kappa_j)\gamma_{1i} \langle \delta \kappa_j(0) \delta \gamma_{1i}
(\theta) \rangle
\biggr],
\nonumber
\\
\label{eq:Iij}
\EEA
for $i=\rm{A,B,C}$.  
In a similar manner, for a four-image system with 
two minima A and C and two saddles B and D with $\kappa_{\textrm{B}}<1$
and $\kappa_{\textrm{D}}<1$, the second moment is given by
\BEA
\langle \eta^2 \rangle&=&\frac{1}{8}
\biggl[I_A+I_B 
-2I_{AB}(\theta_{AB}) + (I_C+I_B)
\nonumber
\\
&-&\!\!2I_{CB}(\theta_{CB})+(I_A+I_D) 
-2I_{AD}(\theta_{AD})
\nonumber
\\
&+&\!\!(I_C+I_D)
-2I_{CD}(\theta_{CD})
\biggr],
\nonumber
\\
\EEA
where $I_i$ and $I_{ij}(\theta)$, $i=\rm{A,B,C,D}$ are given by (\ref{eq:Ii})
and (\ref{eq:Iij}).
Note that we are using coordinates in which $\phi_\theta=0$.

\section{Results}

\subsection{Probability distribution of $\eta$}

\begin{figure*}
\vspace*{1.0cm}
\includegraphics[width=160mm]{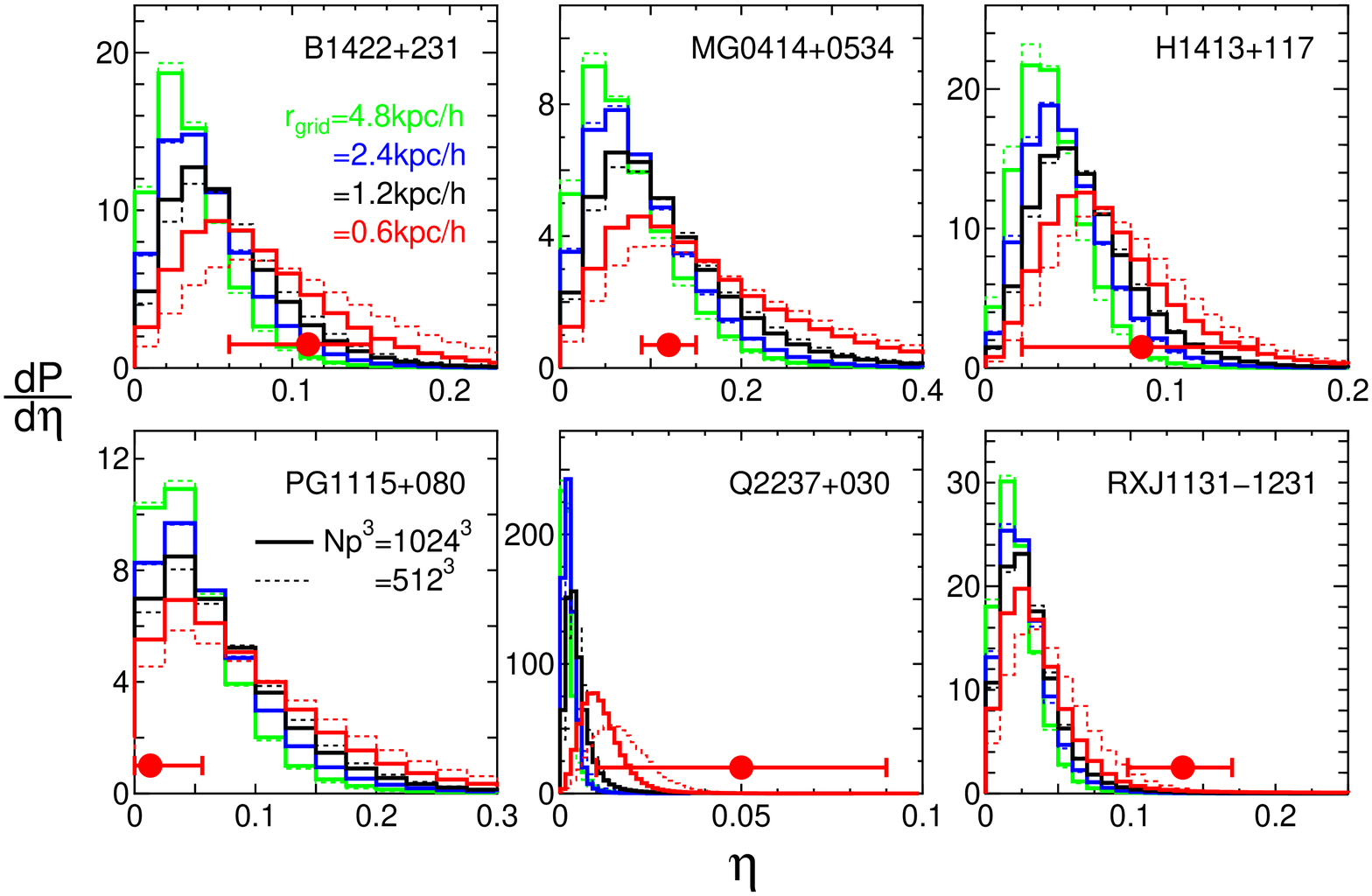}
\caption{
The PDF of $\eta$ for the six lens systems.
The histograms are our simulation results of the $1024^3$ particles (solid
 lines) and $512^3$ particles (dotted lines) with
 the grid
 resolution $r_{\rm grid}= 4.8$(green), $2.4$(blue), $1.2$(black) and
 $0.6 h^{-1}$kpc(red), respectively.
The red filled circles with error bars denote the observational 
 data with the $1 \sigma$ error.  
}
\label{fig_PDF_eta_allsystems}
\vspace*{0.5cm}
\end{figure*}

\begin{table*}
\caption{Mean and rms of $\eta$ for various grid sizes $r_{\rm grid}$}
\begin{tabular}{lccccc}
\hline
  lens system  & $r_{\rm grid}(h^{-1} {\rm kpc})$ & $\langle \eta \rangle
 \pm \Delta \eta$(sim.) &
 $\langle \eta^2 \rangle^{1/2}$(sim.) &
 $\langle \eta^2 \rangle^{1/2}$(CS) & $\langle \eta^2 \rangle^{1/2}$(SK)  \\ 
\hline
  B1422+231  & $0.6$ & $0.081 \pm 0.053$ & $0.097$ & $0.138$ & $0.141$ \\
 $\eta_{\rm obs}=0.11^{+0.04}_{-0.05}$
            & $1.2$ & $0.058 \pm 0.038$ & $0.069$ &  $0.117$ & $0.120$ \\  
            & $2.4$ & $0.047 \pm 0.031$ & $0.056$ &  \\
            & $4.8$ & $0.037 \pm 0.026$ & $0.044$ &  \\
\hline
  MG0414+0534 & $0.6$ & $0.158 \pm 0.103$  & $0.188$ & $0.160$ & $0.182$ \\ 
 $\eta_{\rm obs}=0.12 \pm 0.03$
            & $1.2$ & $0.114 \pm 0.076$  & $0.137$ & $0.129$ & $0.151$ \\ 
            & $2.4$ & $0.091 \pm 0.060$  & $0.109$ &  \\
            & $4.8$ & $0.075 \pm 0.053$  & $0.092$ & \\
\hline
  H1413+117 & $0.6$ & $0.072 \pm 0.044$ & $0.084$ & $0.065$ & $0.073$ \\
 $\eta_{\rm obs}=0.086^{+0.064}_{-0.066}$
            & $1.2$ & $0.055 \pm 0.029$ & $0.062$ & $0.052$ & $0.061$ \\
            & $2.4$ & $0.046 \pm 0.025$ & $0.052$ &  \\  
            & $4.8$ & $0.037 \pm 0.019$ & $0.041$ &  \\
\hline
  PG1115+080 & $0.6$ & $0.090 \pm 0.076$ & $0.118$ & $0.132$ & $0.135$ \\
 $\eta_{\rm obs}=0.013^{+0.043}_{-0.013}$
             & $1.2$ & $0.070 \pm 0.060$ & $0.093$ & $0.115 $ & $0.121$ \\
             & $2.4$ & $0.059 \pm 0.051$ & $0.078$ &  \\
             & $4.8$ & $0.047 \pm 0.042$ & $0.063$ &  \\
\hline
  Q2237+030 & $0.6$ & $0.0120 \pm 0.0111$ & $0.0163$ & $0.011 $ & $0.011$ \\
 $\eta_{\rm obs}=0.05 \pm 0.04$
            & $1.2$ & $0.0053 \pm 0.0049$ & $0.0072$ & $0.002 $ & $0.002$ \\ 
            & $2.4$ & $0.0027 \pm 0.0021$ & $0.0034$ &  \\
            & $4.8$ & $0.0021 \pm 0.0021$ & $0.0030$ &  \\ 
\hline
  RXJ1131-1231 & $0.6$ & $0.040 \pm 0.041$ & $0.058$ & $0.071$ & $0.072$ \\
 $\eta_{\rm obs}=0.136^{+0.034}_{-0.038}$
            & $1.2$ & $0.031 \pm 0.021$ & $0.037$ & $0.064$ & $0.065$ \\ 
            & $2.4$ & $0.027 \pm 0.017$ & $0.032$ &  \\
            & $4.8$ & $0.022 \pm 0.015$ & $0.027$ &  \\
\hline
\end{tabular}
\label{table3}
\end{table*}

In this section, we present the results of our simulation. First we 
show the probability distribution function (PDF) of $\eta$, 
the mean $\langle \eta \rangle$ and the second moment $\langle \eta^2 \rangle$,
 and then we discuss the numerical convergence of our simulation.
Note that the magnification perturbation $\eta$ is 
defined in equation (\ref{eta_def}).
Fig. \ref{fig_PDF_eta_allsystems} shows the PDFs of $\eta$ for the
 six lens systems.
Each colored histogram shows the PDFs with different grid sizes of
 $r_{\rm grid}=4.8h^{-1}$kpc (green), $2.4h^{-1}$kpc (blue),
 $1.2h^{-1}$kpc (black), and $0.6h^{-1}$kpc (red).
The solid and dotted lines
correspond to the different numbers of
$N$-body particles, $N_p^3=1024^3$ (solid) and $512^3$ (dotted).
Each PDF is normalised to unity.
The red filled-circles with horizontal error bars ($1 \sigma$) 
correspond to the observational data. Note that uncertainty of the
primary lens is not taken into account for determining the errors
of $\eta$. On smaller grid scales, the PDFs 
become larger and have a longer tail at large
$\eta$.  As clearly seen in Fig. \ref{fig_PDF_eta_allsystems}, 
our simulation results are consistent with the observed data.

We check the numerical convergence of our simulation by comparing
 the high (low) resolutions of $N_p^3=1024^3 (512^3)$ particles.
The solid (dotted) lines in Fig. \ref{fig_PDF_eta_allsystems} are the
 results for $N_p^3=1024^3 (512^3)$.
As clearly seen in Fig. \ref{fig_PDF_eta_allsystems}, 
the results with  $N_p^3=1024^3$ agree with
those with $512^3$ 
in which  $r_{\rm grid}=4.8,2.4$ and $1.2h^{-1}$kpc for all the
 systems.
However, in the smallest grid size $r_{\rm grid}=0.6h^{-1}$kpc, with
 $N_p^3=512^3$, the distribution is clearly broader than that with 
 $N_p^3=1024^3$ for the all systems.
This is because, for small number of particles, the shot noise
dominates the power spectrum and hence
 the PDF becomes wider.
The contribution of shot noise to the second moment $\langle \eta^2 \rangle$
 can be analytically estimated by replacing the power spectrum $P_\delta(k,r)$
 in equation (\ref{eq:ck}) with $P_\delta(k,r)+1/n$ where $n$ is the number
 density of $N$-body particles and $1/n$ corresponds to the shot
 noise.
For the grid size $r_{\rm grid}=0.6h^{-1}\,$kpc with $N_p^3=512^3$, the shot
 noise dominates the signal.
For the larger grid sizes $r_{\rm grid} \geq 1.2h^{-1}\,$kpc, the small-scale
 $P_\delta$ and the shot noise do not contribute to the integral of
 $\langle \eta^2 \rangle$ in equation (\ref{eq:ck}) due to the window function,
 and hence we can neglect the shot noise even for $N_p^3=512^3$.
In what follows, we use the result of our simulation with 
parameters $N_p=1024^3$ and $r_{\rm grid}=1.2h^{-1}\,$kpc as a fiducial model.

Table \ref{table3} shows the mean and the rms of $\eta$ obtained 
efrom the high resolution simulation ($N_p^3=1024^3$) 
and the analytical calculation using
two types of window function, 
``sharp-k space''(SK) and ``constant shift''(CS) cut.
In the 1st column, we plot the observed $\eta$ and the mean with
au the $1 \sigma$ error, in the 2nd column, the grid sizes from
 $r_{\rm grid} = 0.6h^{-1}$kpc (top) to $4.8h^{-1}$kpc (bottom), 
in the 3rd and 4th columns,  the mean $\eta$ with the  $1 \sigma$ error
 and the rms of $\eta$ obtained from the simulation 
(in calculating these statistical quantities from the simulation,
we use the data up to $\eta=0.5$ to cut the long tail in the
 PDF\footnote{We find that $99.4 \%$ of our samples 
satisfy $\eta \le 0.5$ for all the six lens systems in our fiducial simulation model
 ($N_p^3=1024^3$ with grid size $r_{\rm grid}=1.2h^{-1}$pc).
 If we further add samples that satisfy  $0.5 <\eta \le 1$, 
 $\langle \eta^2 \rangle^{1/2}$ increase by $11 \%$ for MG0414, by
 $2.5 \%$ for PG1115 and by $<0.21 \%$ for the others.
This is because the PDFs of MG0414 and PG1115 have a relatively longer tail
 as shown in Fig.\ref{fig_PDF_eta_allsystems}. Hence there are 
 systematic errors $\sim 10 \%$ associated with the choice of range in $\eta$.}),
in the 5th column, the rms of $\eta$ from the analytical calculation
 discussed in sec.\ref{analytical}.
Our simulation results of $\langle \eta ^2 \rangle^{1/2} $ agree 
with the analytical ones within $\lesssim 40$ percent. 
The SK cut gives systematically smaller values than the CS cut. 
The difference between SK and CS is conspicuous only for MG0414+0534 and
H1413+117 where the comoving size of the lens is large. Otherwise, 
the difference is negligible.
For the larger source redshift, the magnification perturbation $\eta$ 
is larger. For Q2237+030, both the lens redshift and the effective 
Einstein radius (in the
 comoving scale) are the smallest, 
the magnification perturbation is found to be very small.
\subsection{Fitting Formula for PDF of $\eta$}

\begin{figure*}
\vspace*{1.0cm}
\includegraphics[width=130mm]{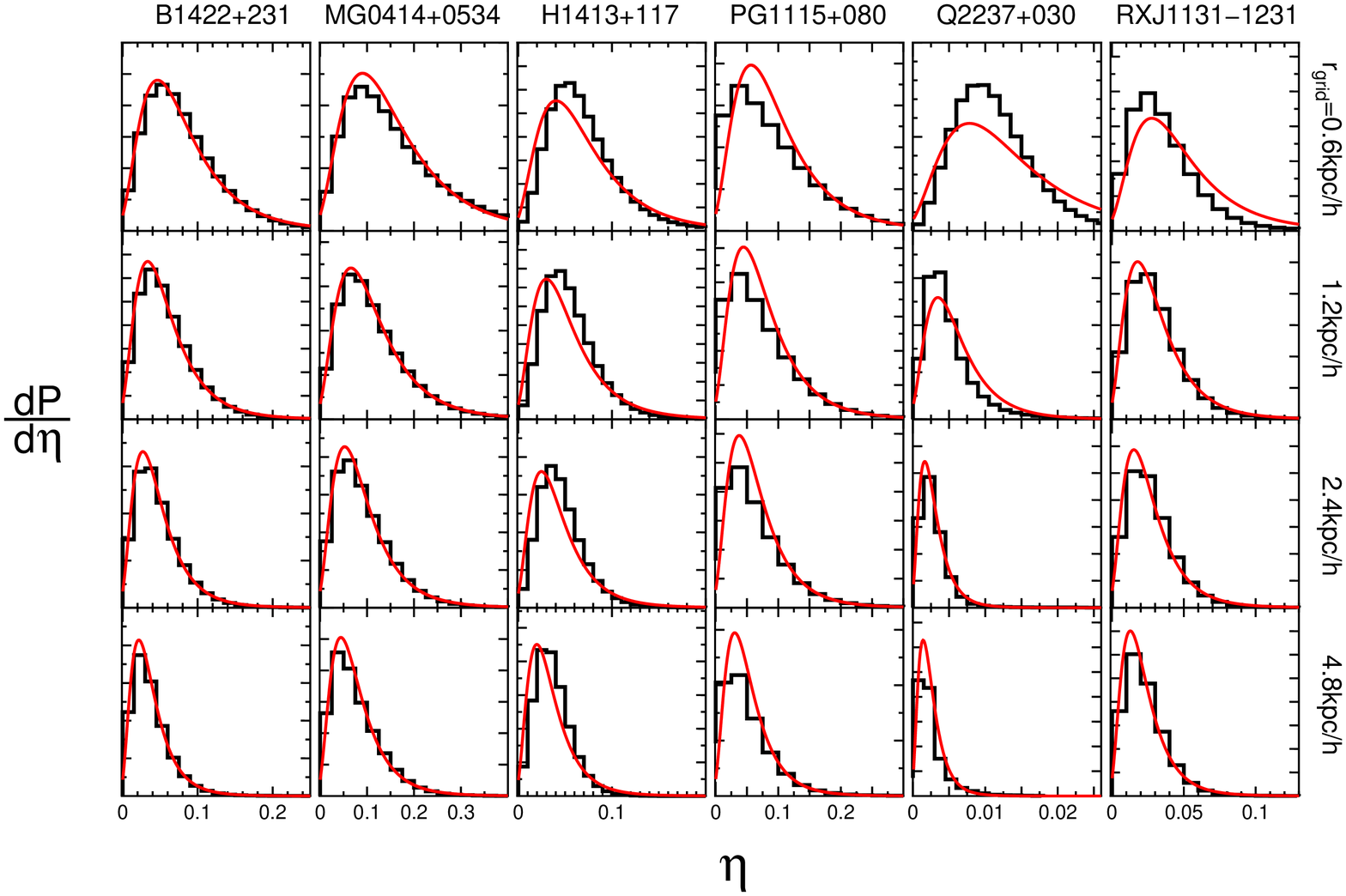}
\caption{
The fitting formula for the PDF of $\eta$ for the six lens systems.
The red curves are the log-normal fitting function, while the black
 histograms are the simulation results for $1024^3$ particles with
 various grid sizes $r_{\rm grid}=0.6-4.8h^{-1}$kpc. 
}
\label{fig_PDF_fitting}
\vspace*{0.5cm}
\end{figure*}

\begin{figure}
\vspace*{1.0cm}
\includegraphics[width=70mm]{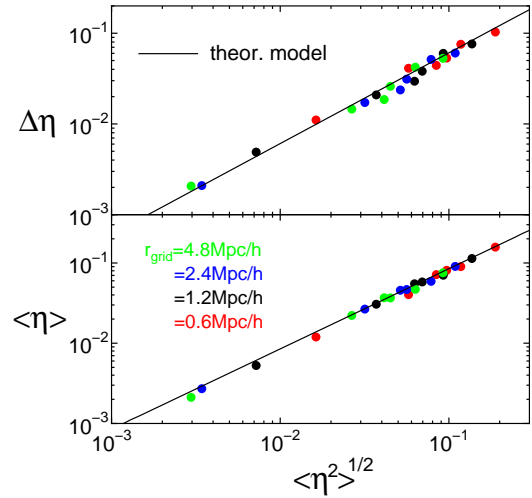}
\caption{
The variance $\Delta \eta$ and the mean $\langle \eta \rangle$
 as a function of the rms $\langle \eta^2 \rangle^{1/2}$.
The colored symbols are the simulation results for the six lens systems
 with various grid sizes.
The solid lines are the theoretical prediction of the log-normal
 PDF.
}
\label{fig_moments_eta}
\vspace*{0.5cm}
\end{figure}

As mentioned in introduction, density fluctuations of dark matter on
 very small scales ($< 10h^{-1}$kpc) are strongly non-Gaussian.
As a result, the perturbed convergence $\delta \kappa$, shear
 $\delta \gamma$ and the magnification perturbation $\eta$ obey non-Gaussian statistics.
Although the analytical formula for the second moment of $\eta$ was derived in
 IT12, it is not sufficient to describe non-Gaussian nature of $\eta$.  
In order to do so, we derive a fitting formula for the
PDF of $\eta$ using our simulated data. 
The PDF is also used to obtain higher order moments of $\eta$ such as 
 the skewness and kurtosis. These statistics contain important
 information about the dark matter clustering on very small scales.

Based on ray-tracing simulations, some authors 
 suggested that the PDFs of convergence and
 shear induced by the large scale
 structures are well approximated by the log-normal distribution
 \citep[e.g.,][]{jain2000,taruya2002,das2006,hilbert2007,takahashi2011}. 
This fact reflects that the one-point distribution function of the matter
 density field is also well described by the log-normal distribution
 \citep[e.g.,][]{coles91,kofman94,kayo01}. As $\eta$ is proportional 
to $\delta \kappa$ and $\delta \gamma$ for $\eta \ll 1$, 
we use the log-normal function for fitting the PDFs of $\eta$,
\BE
  \frac{dP}{d\eta} = N \exp \left[ - \frac{1}{2 \sigma^2}
 \left\{ \ln \left( 1 + \frac{\eta}{\eta_0} \right) - \ln \mu \right\}^2
 \right] \frac{1}{\eta+\eta_0},
\label{pdf_eta}
\EE
where $N$ is a normalisation constant, $\eta_0$ describes a dispersion scale of $\eta$,
 and $\sigma$ and $\mu$ are kept to be constant\footnote{Taruya et al. (2001)
 showed that the PDF of the convergence $\delta \kappa$ is well fitted by
 the following log-normal function,
\BE
 \frac{dP}{d\delta \kappa} = N_\kappa \exp
 \left[ - \frac{1}{2 \sigma_\kappa^2}
 \left\{ \ln \left( 1 + \frac{\delta \kappa}{\left| \delta \kappa_{\rm min}
 \right|} \right) + \frac{\sigma_\kappa^2}{2} \right\}^2
 \right] \frac{1}{\delta \kappa+ \left| \delta \kappa_{\rm min} \right|},
\label{pdf_kappa}
\EE
where $\delta \kappa_{\rm min}$ is the minimum convergence for the empty
 beam \citep[e.g.,][]{jain2000} and $\sigma_\kappa$ is the variance of
 $\delta \kappa$.
Our fitting formula in equation (\ref{pdf_eta}) is based on equation (\ref{pdf_kappa}),
 but simply replace the variable $\delta \kappa$
 to $\eta$ and add a term $\ln \mu$ instead of $\sigma_\kappa^2/2$
 in the exponential to give a better fit at $\eta \sim 0$.}.
We assume that the fitting formula depends only on 
 the second moment $\langle \eta^2 \rangle$.
In fact, the PDF may depend on various factors such as the source
 and lens redshifts, the image positions, and the lens parameters.
However, the most important parameter that characterises the PDF is the
 second moment
 $\langle \eta^2 \rangle$, and hence we simply ignore the other factors.
Then we can calculate $\langle \eta^2 \rangle$ using equation
 (\ref{pdf_eta}) as,
\BE
  \langle \eta^2 \rangle = \int_0^\infty d\eta \frac{dP}{d \eta} \eta^2
 \propto \eta_0^2,
\label{eta_propto}
\EE
and hence $\eta_0 \propto \langle \eta^2 \rangle^{1/2}$.
Then we fit three parameters $\sigma$, $\mu$ and
 $\eta_0/\langle \eta^2 \rangle^{1/2}$ to minimize
the chi-square 
\BE
 \chi^2 = \sum_{\rm lens} \sum_{r_{\rm grid}} \sum_\eta
 \frac{\left( dP_{\rm model}/d\eta - 
 dP_{\rm sim}/d\eta \right)^2}{\left( dP_{\rm model}/d\eta \right)^2}  
 \delta \eta
\EE
where $dP_{\rm model}/d\eta$ and $dP_{\rm sim}/d\eta$ are the PDFs for the
 log-normal model and the simulation results, respectively, and $\delta \eta$
 is the bin-width in the simulation results of the PDFs.
The summation is performed over all the six 
lens systems, the four grid sizes $r_{\rm grid}=0.6-4.8h^{-1}\,$kpc, 
and $\eta$. Then, we find the best-fit parameters as, 
\BE
 \mu=4.10, ~\sigma^2=0.279, ~\eta_0=0.228 \langle \eta^2 \rangle^{1/2},
\label{pdf_eta_params}
\EE 
where the second moment $\langle \eta^2 \rangle$ is obtained from the
 simulation (see Table \ref{table3}).
Fig. \ref{fig_PDF_fitting} shows our best-fit model (red curves) 
and the simulation results (black histograms).
The best-fit model is in agreement with the simulation results within an
error of $\lesssim 30 \%$.

Using the log-normal PDF in equation (\ref{pdf_eta}), one can easily verify that
the $n$-th moment of $\eta$, $\langle \eta^n \rangle$, is proportional to $\eta_0^n$.
Then, equation (\ref{eta_propto}) implies that the mean $\langle \eta
 \rangle$ and the standard deviation $\Delta \eta$ are
 also proportional to $\eta_0$ and the rms of $\eta$, $\langle \eta^2
 \rangle^{1/2}$. The relations between these variables can be 
seen in Fig. \ref{fig_moments_eta}.
The horizontal axis denotes $\langle \eta^2 \rangle^{1/2}$ and
the vertical axis shows the standard deviation $\Delta \eta$ and the
 mean $\langle \eta \rangle$, respectively.
The colored dots are the simulation results for the six lens systems 
with various grid sizes.
The solid straight lines correspond to 
the theoretical values based on the log-normal PDF.
Equation (\ref{pdf_eta}) and the best-fit parameters in 
equation (\ref{pdf_eta_params}) lead to
\BE
 \Delta \eta = 0.608 \langle \eta^2 \rangle^{1/2}, ~~
 \langle \eta \rangle = 0.850 \langle \eta^2 \rangle^{1/2}.
\EE
As shown in Fig. \ref{fig_moments_eta}, these best-fit 
parameters give good approximations of $\Delta \eta$
 and $\langle \eta \rangle$ as a function of $\langle \eta^2
 \rangle^{1/2}$, which can be estimated using the analytical 
formulae in section 6. As these values do not depend on 
complex parameters of each lens system, it may be considered as 
the ``universal'' relation which is applicable to all quadruple 
lens systems.

\subsection{PDF of magnification contrast}

\begin{figure*}
\vspace*{1.0cm}
\includegraphics[width=160mm]{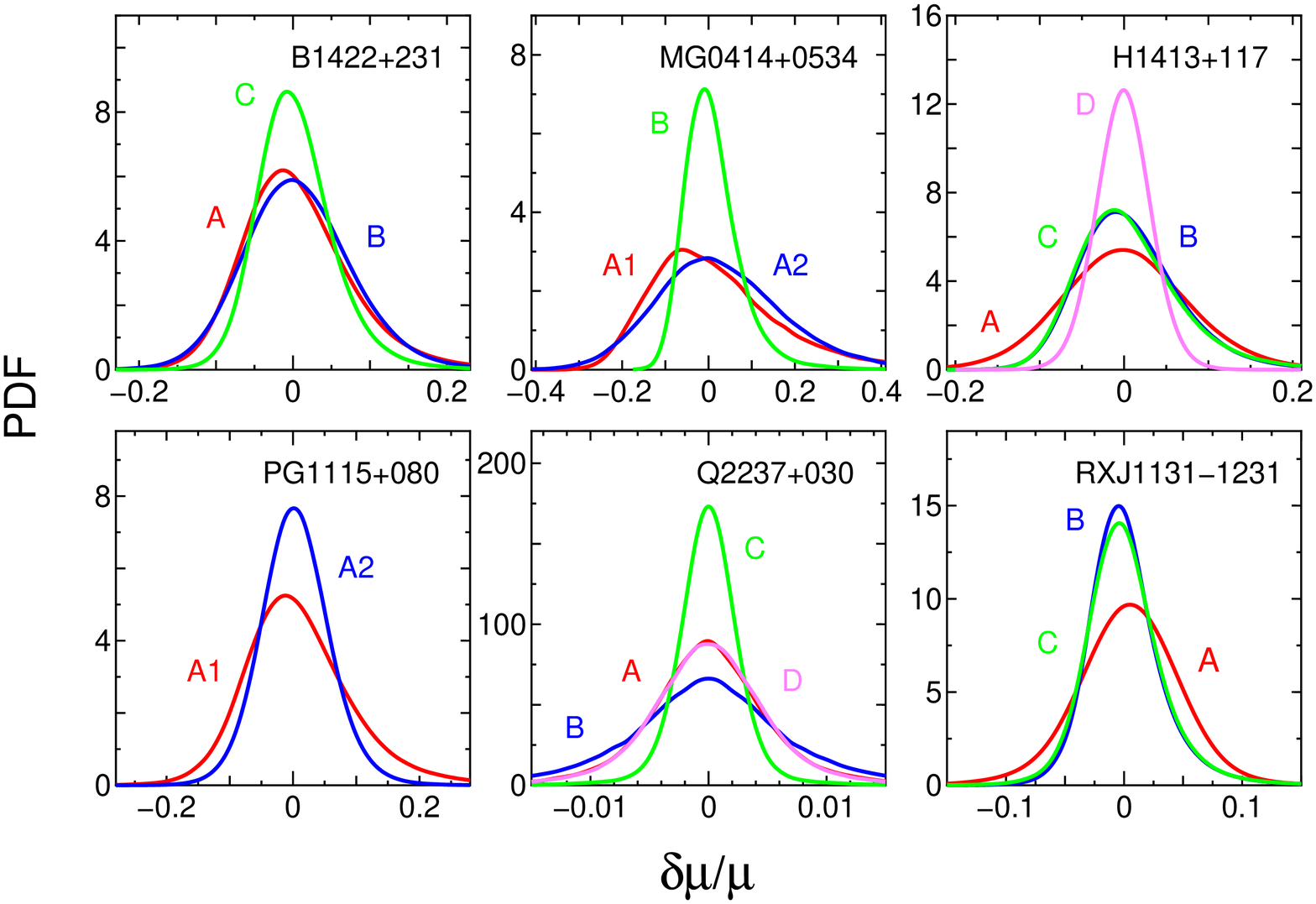}
\caption{
The PDF of the magnification contrast $\delta \mu/\mu$ of each image.
Each colored curve corresponds to the PDF of each image A,B,C$\cdots$.
The simulation results are for $1024^3$ particles with the grid size $r_{\rm grid}= 1.2 h^{-1}$kpc.
}
\label{fig_PDF_distmu_allsystems}
\vspace*{0.5cm}
\end{figure*}

\begin{table}
\caption{Standard deviation and skewness of $\delta \mu/\mu$}
\begin{tabular}{lcccc}
\hline
  lens system  & image & parity & deviation $\sigma$ & skewness $S$  \\ 
\hline
  B1422+231 & A & $+$ & 0.074 & 0.86  \\
            & B & $-$ & 0.072 & 0.31  \\
            & C & $+$ & 0.052 & 0.82  \\
\hline
  MG0414+0534 & A1& $+$ & 0.14 & 0.73  \\ 
              & A2& $-$ & 0.15 & 0.34  \\ 
              & B & $+$ & 0.067 & 1.3  \\
\hline
  H1413+117 & A & $-$ & 0.078 & 0.18  \\
            & B & $+$ &  0.062 & 0.91  \\
            & C & $+$ & 0.064 & 1.1  \\
            & D & $-$ & 0.032 & 0.042  \\
\hline
  PG1115+080 & A1 & $+$ & 0.089 & 0.83  \\
             & A2 & $-$ & 0.056 & 0.25  \\ 
\hline
  Q2237+030 & A & $+$ & 0.0067 & 0.62  \\
            & B & $+$ & 0.010 & 0.54  \\
            & C & $-$ & 0.0030 & 0.10  \\
            & D & $-$ & 0.0065 & 0.20  \\
\hline
  RXJ1131-1231 & A & $-$ & 0.045 & -0.0071  \\
               & B & $+$ & 0.033 & 1.3  \\
               & C & $+$ & 0.035 & 1.1  \\
\hline
\end{tabular}
\label{table4}
\end{table}
This section shows the simulation results for the 
PDFs of magnification
contrast $\delta \mu/\mu$ for each
image in the six lens systems (Fig. \ref{fig_PDF_distmu_allsystems}).
We find that the changes in the fluxes are typically $\sim 1-10\%$, 
and the magnification contrast $\delta \mu/\mu$ is larger for
systems with a higher source redshift.
For Q2237, the nearest lens at $z_L=0.04$ 
is exceptionally small in comparison with the other lense systems, 
the flux change is quite small
because the relevant comoving size of the lens 
is significantly small in comparison with the other lens systems.
The PDF of $\delta \mu/\mu$ is broader for brighter images, 
because the magnification perturbation is proportional to the
 unperturbed magnification due to the primary lens,
 $\delta \mu_{\rm i}/\mu_{\rm i} \propto \mu_{\rm i}$, which can be
 shown by using equations (\ref{relative_mu}) and (\ref{eta_approx}).
The standard deviations of $\delta \mu/\mu$,
 $\sigma^2 = \langle \left( \delta \mu / \mu - \langle \delta \mu / \mu
 \rangle \right)^2 \rangle$, are listed in the 3rd column in
 Table \ref{table4}.
The means of $\delta \mu/\mu$ are consistent to be zero.
Table \ref{table4} also show the skewness, $S = \langle ( {\delta \mu}/{\mu}
 - \langle {\delta \mu}/{\mu} \rangle )^3 \rangle / \sigma^{3/2}$,
 in the 4th column.
If the PDF of $\delta \mu/\mu$ obeys Gaussian statistics, the skewness
vanishes.
In calculating the above moments, we use the range
 $\left| \delta \mu / \mu \right| <0.5$.
As shown in Table \ref{table4}, the skewness for 
images with a positive parity is systematically
larger than those with a negative parity. This reflects the fact that 
images with a positive parity tend to be magnified by a halo while those with a
negative parity are either magnified or demagnified by a halo at the
linear regime depending on the positions of the perturbers. 
Although the number of halos are smaller than that of
voids, the amplitude of the density contrast due to 
halos can be higher than that due to voids. 
Therefore, we have a positive skewness for images with a
positive parity. On the other hand, for images with a negative parity, halos can
demagnify the image. Hence, the skewness becomes small in comparison
with the images with a positive parity. If the perturber is a void, then 
images with a positive parity are demagnified and those with a negative 
parity are either magnified or demagnified. If the origin of the flux
anomaly is the weak lensing effect due to the line-of-sight structures
rather than subhalos, we would observe these patterns in flux changes if
perturbed by voids. However, the number of our sample is not enough to
confirm such cases.

\subsection{Mass scales of intervening structures}

\begin{figure*}
\vspace*{1.0cm}
\includegraphics[width=175mm]{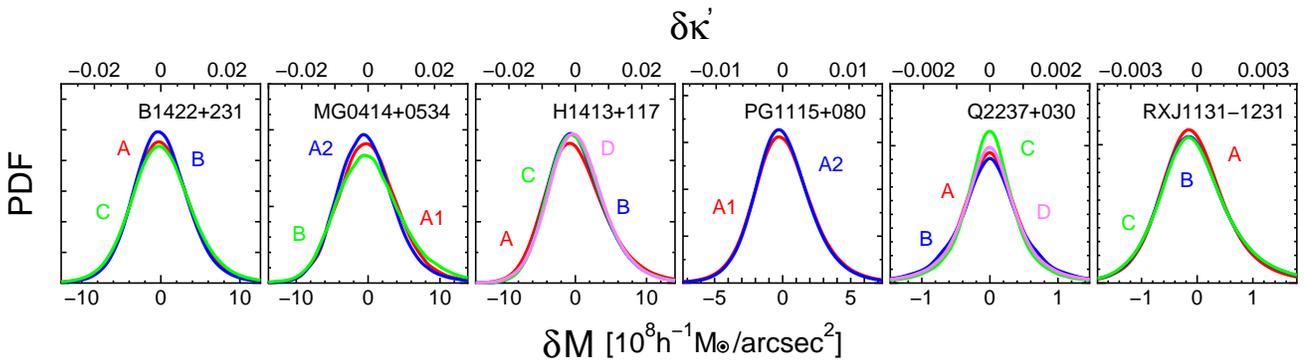}
\caption{
The PDFs of the approximated surface mass density of intervening structures
 $\delta M$ per $\rm{arcsec}^2$, in unit of $10^8 M_\odot $. 
We assume that $D_{j}=D_{jS}$ in equation (\ref{mass_dens}) where
the contribution becomes maximal.
Each colored curve corresponds to $\delta M$ along the
line-of-sight to each image.
The upper horizontal axes also show the corresponding convergence
 perturbation.
The simulation results are for $1024^3$ particles with a grid
 size $r_{\rm grid}= 1.2 h^{-1}$kpc.
}
\label{fig_pdf_kappa}
\vspace*{0.5cm}
\end{figure*}

In this subsection, we estimate the mass scales
of intervening structures in 6 quasar-galaxy lens systems.
In Fig.\ref{fig_pdf_kappa}, we plot the PDFs
 of surface mass $\delta M$ due to intervening structures 
per ${\rm{arcsec}}^{2}$ for all the lens systems.
The horizontal axis is the intervening surface mass 
(in unit of $10^8 M_\odot $) per ${\rm arcsec}^2$.
The surface mass density is obtained from the convergence perturbation
 $\delta \kappa^\prime$ using the relation (e.g. Schneider et al. 1992),
 $\delta M = \delta \kappa^\prime \Sigma_{\rm cr} D_j^2$, where 
 $\Sigma_{\rm cr}$ is the critical mass density and $D_j$ is the angular
 diameter distance to the intervening strcture on the j-th lens plane
 (see Fig.\ref{fig_lens_planes2}).
Then we have,
\BE
 \delta M = 1.17 \times 10^9 h^{-1} M_{\odot} {\rm arcsec}^{-2}
 \left( \frac{\delta \kappa^\prime}{0.01} \right) 
 \frac{H_0 D_j D_S}{D_{jS}},
\label{mass_dens}
\EE 
where $D_{jS}$ and $D_S$ are the angular diameter distances from the
 j-th lens plane and the observer to the source, respectively.
The upper horizontal axes of Fig.\ref{fig_pdf_kappa} show the
 corresponding convergence perturbations $\delta \kappa^\prime$.
For obtaining the convergence perturbations $\delta \kappa^\prime$, we 
subtract the mean convergence within the effective 
Einstein radius of the primary lens (see sec. \ref{5.2}). 
In order to estimate the surface mass density, we assume 
that $D_{j}=D_{jS}$ in equation (\ref{mass_dens}) where
the contribution becomes maximal. The results 
simply scale as $\delta M \propto D_j/D_{jS}$ if 
$D_{j} \neq D_{jS}$. Therefore, the obtained values should be
interpreted as the upper limits.

As shown in Fig.\ref{fig_pdf_kappa}, {\it{the obtained 
surface mass densities are $O[10^{7-8}] M_\odot {\rm arcsec}^{-2}$
and the contribution from negative masses (voids)
are comparable to that from positive masses (halos)}}.
As the distance to the source increases (decreases), or the
comoving size of the lens increases (decreases), 
the projected mass of intervening structures gets larger (smaller). 
The result clearly shows that the lens systems with a 
high redshift source or a massive lens are 
much affected by the line-of-sight structures. 

Although the convergence perturbations are independent on angular positions
 due to the spatial homogeneity of perturbations,
 there are tiny differences between the PDFs within one single lens system.
This is because that 1) we subtract the mean convergence perturbation around
 the lens center and hence $\delta \kappa^\prime$ depends on the
 angular positions with respect to the 
lens center, and 2) we also impose various
 conditions (see sec.\ref{5.2}).  

In our model, the effect of structures whose sizes are equal to or larger than
the size of the primary lens is taken into account as part of
the primary lens (e.g, external shear, constant convergence). Therefore,
the angular sizes of the ``background'' intervening structures are smaller than
the size of the primary lens, and hence their mass scales (void or halo)
are $\lesssim 10^8 M_\odot$.

\subsection{Comparison with Observed Flux Ratios}

\begin{figure*}
\vspace*{1.0cm}
\includegraphics[width=150mm]{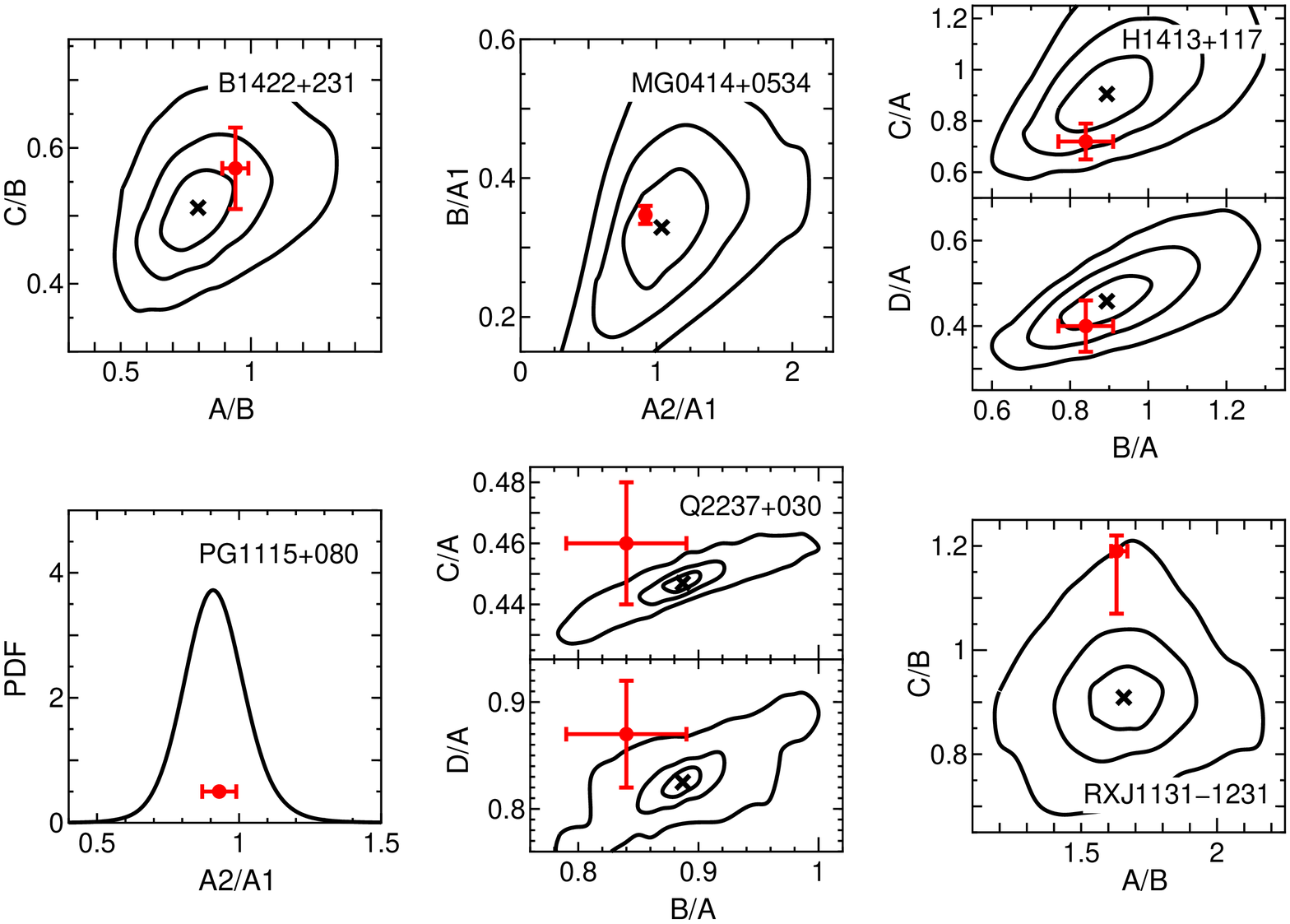}
\caption{
Flux ratios of lensed images for six
 MIR lens systems. 
The horizontal and vertical axes are
 the flux ratios in our simulation except for
 PG1115+080 where the PDF as a
 function of the
 flux ratio A1/A2 is plotted, the black crosses
 show the best-fit values that describe the primary lens models.
The black contours around the crosses correspond to $1 \sigma, 2 \sigma$,
 and $3 \sigma$ confidence regions due to intervening structures,
 and the red filled circles are the measured
 results with the $1 \sigma$ error.
}
\label{fig_flux_ratios}
\vspace*{0.5cm}
\end{figure*}

In this subsection, we compare our
 theoretically predicted flux ratios for $1024^3$ particles 
with $r_{\rm grid}=1.2h^{-1}\,$kpc to the
 observed values for the 6 MIR
 lens systems (Fig. \ref{fig_flux_ratios}). 
The horizontal and vertical axes are the flux ratios except for
 PG1115+080 where the PDF as a function of the
 flux ratio A1/A2 is plotted, the black crosses
 show the best-fit values that describe the primary lens models,
 the black contours around the crosses correspond to $1 \sigma, 2 \sigma$,
 and $3 \sigma$ confidence regions due to intervening structures,
 and the red filled-circles with bars are the measured
 results with the $1 \sigma$ observational error.
The predicted flux ratios are in good agreement with the 
observed values at $1 \sim 2 \sigma$ level. 
Thus, the intervening structures can provide sufficient
magnification perturbations in 
order to explain the anomalies in the flux-ratios.

\subsection{Systematics}
\label{systematics}

Finally, we comment on some systematics in our simulations.
In our procedure, we impose various conditions and assumptions on
 our simulations in evaluating the magnification perturbations.
In this section, we investigate the effects of these
 conditions on our results.
We present the results only for B1422+231 with $r_{\rm grid}=1.2h^{-1}$kpc,
but we expect that similar results hold for the other lens systems.

In our simulations, 
we impose a condition that massive line-of-sight halos do not reside 
around a lens (see section \ref{5.2}) since the magnification perturbation 
would be too large if they are included.
Furthermore, we impose the following two conditions: (i) maximum
perturbed convergence within a circle with an angular radius
 $10^{\prime \prime}$ centered at the lens center is less than $0.5$, and
 (ii) perturbed convergence and shear within a circle with the effective
 Einstein radius of the primary lens centered at the lens center 
should be less than those of the primary lens 
(see procedure 1) in section \ref{5.2}). 
If we do not impose these conditions, 
the mean and rms of $\eta$ systematically change $\sim 1\%$ at most.
Similarly, we also impose that the relative shifts in positions of
lensed images due to the line-of-sight structures 
should be less than $0.0042^{\prime \prime}$.
Without this condition, the rms of $\eta$ increases by about $60 \%$
 ($\langle \eta^2 \rangle^{1/2}=0.069$ increases to $0.11$).
Hence, the constraint on angular shifts of lensed images significantly 
influences the magnification perturbations.  

In section \ref{5.2}, procedure 2), we added the perturbed convergence
 $\delta \kappa$ and shear $\delta \gamma_{1,2}$ 
 to $\kappa$ and $\gamma_{1,2}$ in the primary lens.
Then, we subtracted the mean perturbed convergence
 $\delta \bar{\kappa}$ and shear $\delta \bar{\gamma}_{1,2}$ around the
 lens from $\delta \kappa$ and $\delta \gamma_{1,2}$.
This is because that the primary lens model already 
includes the external shear and constant density.  
Similarly, we also subtract the mean perturbed deflection-angles for the
 same reason. To calculate the mean of perturbed quantities, we average
them over a region within a circle with
an effective Einstein radius $\theta_E$ 
of the primary lens centered at the primary
 lens center. However, it is not simple to divide a complex lens system into
 the ``external'' and ``internal'' parts. For instance, actual lenses
 generally have a non-zero ellipticity, and the light-ray paths around a lens
 center are not so simple as discussed in
 section \ref{3.2}. To check how these uncertainties would affect our result, 
 we calculate mean quantities over regions with circle radii, 
$2\theta_E$ and $3 \theta_E$. 
Then the rms's of magnification perturbation $\eta$ 
are found to be $\langle \eta^2 \rangle^{1/2}=0.069, 0.077$ and
 $0.092$ for $\theta_{\rm E}$, $2 \theta_{\rm E}$ and $3 \theta_{\rm E}$, 
 respectively. Hence, such ambiguity would change the magnification
 perturbations by $\sim 20$ percent.

It is also important to comment on the mass-sheet degeneracy in the lens model. 
The flux ratios for the primary lens (without perturbations) 
are invariant under the transformation, $1-\kappa_{\rm i}
 \rightarrow \lambda (1-\kappa_{\rm i})$, $\gamma_{{\rm i} 1,2} \rightarrow
 \lambda \gamma_{{\rm i} 1,2}$ and $\beta \rightarrow \lambda \beta$
 where $\lambda$ is a constant \citep{falco85}. 
Under this transformation, the magnification perturbation is transformed
 as $\eta \rightarrow \lambda^{-1} \eta$ and $\delta \mu_{\rm i}
 / \mu_{\rm i} \rightarrow \lambda^{-1} \delta \mu_{\rm i}/\mu_{\rm i}$.
As the typical amplitude of the background shear on a scale of an Einstein
radius $\theta_E=1''$ (the typical size of a lens in our samples) is
$\gamma \lesssim 0.1$ (see Fig. \ref{fig_PDF_kappam}), 
the degeneracy affects $\eta$ by $\lesssim 10$ percent. 

In our ray-tracing simulation, we assumed the Born approximation which has some limitations and systematics. Under the approximation, we cannot investigate the strong lensing effects by the intervening objects such as image-position shifts and strong image deformations. Especially, the image-position shifts are very important, since we can separate the contributions from the main lens and the intervening lenses by fitting the image positions and then we can correctly estimate the magnification perturbations induced by the intervening objects. As a result, our conclusions would strongly depend on this approximation and hence we are planning to re-calculate the simulations without the Born approximation.

\section{Conclusion and Discussion}

We have investigated the weak lensing effects of 
line-of-sight structures in quadruple lens systems.
We have run ray-tracing simulations to evaluate the magnification
perturbations of quadruply lensed images.
The magnification ratios typically change by $O(10\%)$ 
when the shifts of relative angular positions of quadruple lensed images 
are constrained to $<0.004''$. We have found that the probability of
having such small shifts is extremely small for MG0414+0534 and
H1413+117. Therefore, it is likely that object X in these systems are residing
in the intergalactic space rather than the neighbourhood of lensing galaxies.
The constrained amplitudes of projected density 
perturbations due to line-of-sight structures are 
$O(10^8)\,\ms/\textrm{arcsec}^2$. These results are consistent with our new 
analytical estimate that intervening structures can 
perturb the magnifications of lensed images to explain 
the observed anomalies in the flux ratios of quadruple lenses.
We have found that the constrained mean amplitudes of 
projected density perturbations in the line-of-sight are negative, which
suggests that the lensed images are perturbed 
by minivoids and minihalos in underdense regions and approximately 50
 percent of systems are perturbed by voids rather than haloes. Thus the
 role of voids or underdense regions in the line-of-sight is important
 in order to understand the origin of anomaly in the flux ratios.   
We have derived a new fitting formula for evaluating the PDF 
of magnification perturbation, which will be useful for constraining 
models.

In this work, we have incorporated constraints from 
astrometric shifts which have been neglected in literature;
1$\sigma $ errors in the positions of lensed images 
in optical/NIR observations by HST
are used for constraining the possible contribution from line-of-sight
structures. However, these values should be considered as upper
limits. If we could observe our MIR lenses with better accuracy in the 
positions, the errors might become smaller. 
Then, we should observe anomaly in positions of lensed images 
if we use smooth potential for modeling the primary lenses \citep{sluse2012}. 
Therefore, improving accuracy in the positions is 
necessary for obtaining the lower limits
for magnification perturbations. We have considered the effect of 
reparametrisation of the primary lens model. As we have
shown in appendix A, this cannot significantly improve the shift of
positions of lensed images. However, it may improve the fit of positions
at the order $O(\varepsilon)$. In order to check this effect, we need to
perform full Monte-Carlo simulations in which the model parameters are
fit together with the perturbations simultaneously. For estimating the
magnification perturbation, we have considered two
types of filtering, namely ``sharp k-space cut'' and ``constant-shift
cut'' for filtering perturbations that allow shifts of $\varepsilon$ for
each image. We have found that the predicted values of
magnification perturbation $\eta$ 
are consistent with the numerical results within an error of $\lesssim
30 \%$ regardless of types of filtering. In order to filter out
perturbations that give the best-fit model more accurately, 
we need to assess the accuracy of filtering functions 
more carefully using Monte-Carlo simulations. 

In our simulations, we did not include massive line-of-sight haloes
 which induce large deflection angles ($>0.0042^{\prime \prime}$). 
However, some larger deflection cases may also fit the positions and fluxes
 by reparametrizing the primary lens. As a result, we emphasize again that
 our results of magnification perturbations would be a lower limit because
 it does not include larger deflection cases.

In our simulations, we can resolve halos with a mass of $\sim 10^5\, \ms$.
The corresponding spatial scale is $O[1]\,$kpc. 
However, contribution from much smaller haloes may not be negligible.
We expect a boost in magnification perturbations until the size of
halo reaches the source size $O[1]\,$pc in the MIR band. 
Therefore, we need to resolve much smaller haloes.  
$N$-body simulations with increased number of 
particles are necessary for accomplishing this purpose.

For simplicity, we did not 
incorporate baryonic processes such as star formation, gas cooling,
 radiation feedback in our simulations.
Although such baryonic effects 
may enhance density fluctuations at small 
scales ($< 1h^{-1}$Mpc) \citep[e.g.,][]{jing06,vandaalen2011}, we expect
that such effects play relatively minor roles in haloes 
with a mass of $M<10^7 M_\odot$ since it is difficult to maintain star 
forming activity in such shallow potential wells.
Indeed, the relevant mass scales of intervening structures are
 equal to or less than $O[10^8] M_\odot$. 
Therefore, the effect of baryon cooling may be 
limited except for the central regions of low-mass haloes, though
more elaborate analysis should be done to confirm this point.

In order to distinguish between subhalos and the line-of-sight 
mini-structures (halos or voids), 
we need to detect the extended source effect
\citep{koopmans2005, inoue2005a}. Submillimeter imaging 
of anomalous quadruple lensing with angular 
resolution of $<0.1''$ using ALMA will be very 
useful for measuring the mass and
redshift of perturbing objects \citep{koopmans2005, inoue2005b}.  
We will address these issues in future work.  

\section{Acknowledgments}
We thank Takahiro Nishimichi and Takashi Hamana for kindly providing us 
 the numerical codes.
This work was supported in part by JSPS Grant-in-Aid for
Scientific Research (B) (No. 25287062) ``Probing the origin
of primordial mini-halos via gravitational lensing phenomena'', 
 and by Hirosaki University Grant for
Exploratory Research by Young Scientists.
Numerical computations were carried out on SR16000 at YITP in Kyoto
University and Cray XT4 at Center for Computational Astrophysics,
CfCA, of National Astronomical Observatory of Japan.
\vspace{1cm}

\appendix

\section{Reparametrisation of nearly circular SIE models}
In this appendix, we show that reparametrisation of 
model parameters of the best-fit SIE model is impossible if
a certain position of a lensed image is significantly 
perturbed while the others remain 
within the observed errors provided that the eccentricity 
of the SIE is sufficiently small. For simplicity, we assume that the
effective Einstein radius of the best-fit 
SIE is $1\,\textrm{arcsec}$
and all the $1 \sigma$ errors of observed positions 
of lensed images and the centroid
of the primary lensing galaxy are equivalent to a constant 
$\varepsilon \ll 1\,$arcsec. 

The unperturbed lens equation for the $j$-th lensed image is then
\BE
 \BETA = \THE_j - \ALP (\THE_j; \largec),
\EE
where $\largec$ denotes a set of parameters of an SIE, namely, the angle of 
the minor axis of ellipitic contours of the projected surface mass 
density $\theta_e$ with respect to the horizontal axis, 
the ellipticity $e \ll 1$, and the effective Einstein radius $b'$.   
$\BETA, \THE_j$ are the positions of a point source and the best-fit 
$j$-th lensed image and $\ALP$ is the reduced deflection angle. All the
positions of lensed images $\THE_j,$ for $j=1,2,3,4$ fall within 
a radius of $\varepsilon$ centred at the observed positions $\tilde {\THE}_j$.

Suppose that the $j$-th lensed image placed at the neighbourhood of 
a critical curve is significantly perturbed so that the position
of the $j$-th lensed image is shifted from $\THE_j$ to $\THE_j+\delta
\THE_j$ where $\delta \theta_j \gg \varepsilon$ whereas other lensed
images are not perturbed. Then the perturbed lensed equation is 
\BE
 \BETA+ \delta \BETA  = \THE_j + \delta \THE_j - 
\ALP (\THE_j+\delta \THE_j; \largec) - \delta \ALP(\THE_j+\delta \THE_j) ,
\EE
where $\delta \BETA$ and $\delta \ALP$ denote 
the shift of the source due to reparametrisation 
and the reduced deflection angle with $O(|\delta \ALP(\THE_j)|)\gg
\varepsilon$, respectively. In order to 
fit the observed position $\tilde{\THE}_j$, we reparametrise the
parameters $\largec$ to $\largec + \delta \tilde{\largec}$. Then the image 
position changes by $\delta \tilde{\THE_j} $.  Since the lensed images are placed at the neighbourhood of 
the critical curve, the shift of the source position 
$\delta \beta$ should be much smaller
than $\delta \theta_j$ and $\delta \alpha$. Then, the best-fit reparametrised
model should satisfy 
\BE
\biggl| \f{\del \ALP}{\del \THE_i}\delta \tilde{\THE_i}
+ \f{\del \ALP}{\del \largec}\delta \tilde{\largec}\biggr|\le \varepsilon,
\label{eq:condition}
\EE 
and the perturbation of positions 
satisfy $\delta \tilde \theta_j \gg \delta \tilde \theta_i$ for $i \neq j$.
The order of positions are 
$O(\delta \tilde \theta_i)=O(|\delta \tilde \THE_i|)=O(\varepsilon)$
for $i \neq j$. We have also 
\BE
\f{\del \ALP}{\del \THE_j}\delta \tilde{\THE_j}
+ \f{\del \ALP}{\del \largec}\delta \tilde{\largec}\approx \delta \tilde{\THE_j}.
\label{eq:condition2}
\EE
If the maximum orders of components 
of $\del \ALP /\del \THE_i $ and $\del \ALP/\del \largec$ are $O(1)$, then we cannot find any reparametrisation
that satisfies equation (\ref{eq:condition}), 
since we have $O(\delta \tilde{\largec})=O(\delta \tilde{\theta_j})\gg
O(\delta \tilde{\theta}_i)=O(\varepsilon)$. 

First, we estimate the order of components of matrices 
$\del \ALP /\del \THE_i $.
If the positions of lensed images in circular
coordinates are represented as 
$\THE=(\theta \cos{\phi}, \theta \sin{\phi})$, then
the caustic for $b'=1$ is given by 
\BEA
\beta_1
&=&\f{\sqrt{q}}{\sqrt{\cos^2{\phi}+q^2 \sin^2{\phi}}}\cos{\phi}
\nonumber
\\
&-&
\f{\sqrt{q}}{\sqrt{1-q^2}}\biggl[\sinh^{-1}
\biggl(\f{\sqrt{1-q^2}}{\sqrt{q}} \cos{\phi} \biggr) \biggr] 
\label{eq:a4}
\\
\beta_2
&=&\f{\sqrt{q}}{\sqrt{\cos^2{\phi}+q^2 \sin^2{\phi}}}\sin{\phi}
\nonumber
\\
&-&
\f{\sqrt{q}}{\sqrt{1-q^2}}\biggl[\sin^{-1}
\biggl(\sqrt{1-q^2} \sin{\phi} \biggr) \biggr], 
\label{eq:a5}
\EEA 
where $q=1-e$. For $e \ll 1$, equations (\ref{eq:a4}) and (\ref{eq:a5}) 
can be reduced to
\BE
(\beta_1,\beta_2) \approx \f{2 e}{3}(-\cos^3{\phi},\sin^3{\phi}).
\EE
Therefore, the width of the asteroid-shaped caustic is $\Delta \beta=
\sqrt{2}e/3 \ll 1$. As the source is placed inside the caustic, we have
$\beta \approx 0$ and $\ALP \approx \THE_i$. Thus $\del \ALP /\del
\THE_i \approx \boldsymbol{1} $ and $O(\del \ALP /\del
\THE_i)=1$.  
 
Second, we estimate the order of the components of a matrix 
$\del \ALP/\del \largec$, where $\largec=(b',e,\theta_e)$.
As the reduced deflection angle $\ALP$ is proportional to $b'$,  
we have $\del \ALP/\del b' \approx \THE_i$. If we assume that the 
lensed images are sufficiently magnified, then we have $\theta_i \approx
1$ if $b'\approx 1$. Thus $|\del \ALP/\del b'|\approx 1$. In a
similar manner, we obtain $|\del \ALP/\del \theta_e|=|\ALP|=\theta_i
\approx 1$.
Assuming $e \ll 1$ and $b' \approx 1$, the reduced deflection angle can
be approximated as 
\BEA
\alpha_1 &=& \cos{\phi}-\f{e (\cos^3 \phi+3 \cos \phi \sin^2{\phi})}{6}
\\
\alpha_2 &=& \sin{\phi}-\f{e(\sin^3 \phi+3 \,e\sin \phi \cos^2{\phi})}{6},
\EEA
where the position of a lensed image is $\THE=(\theta \cos{\phi}, \theta
\sin{\phi})$. Therefore, $|\del \ALP/\del e|\approx \sqrt{5-3 \cos{4
\phi}}/(6 \sqrt{2})<1/3$. Thus, we have $O(|\del \ALP/\del e|) < 1$
and the maximum order of the components of 
$\del \ALP/\del \largec$ is $1$.

\section{TSC smoothing scheme }
In this appendix, we review the 
``Triangular-Shape-Cloud'' (TSC) scheme in order to 
calculate a smooth distribution from particles.
We derive the window function of the TSC scheme
in the Fourier space in two-dimensional systems.

\begin{figure*}
\vspace*{1.0cm}
\includegraphics[width=110mm]{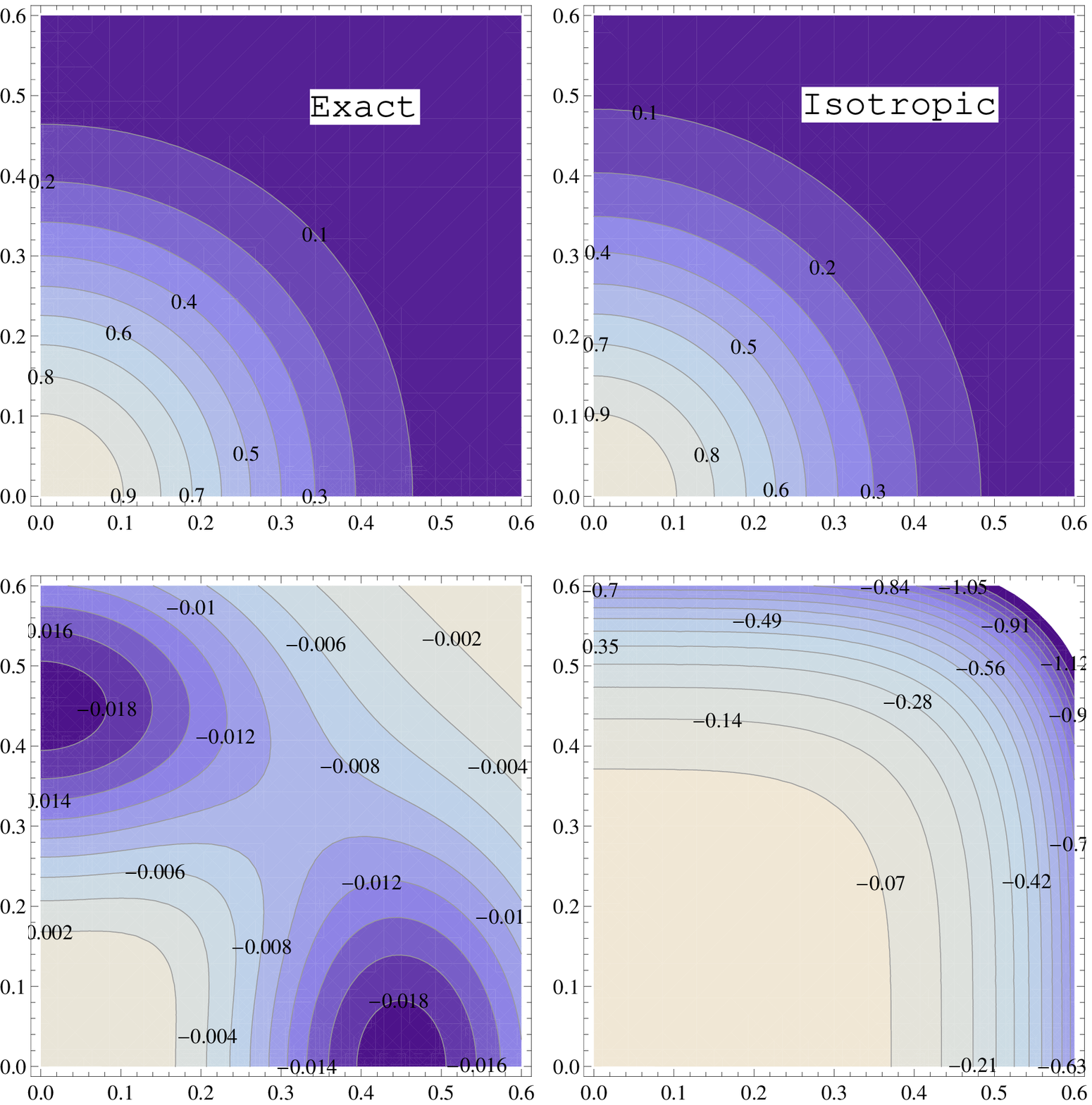}
\caption{Contour plots of the squared TSC window function $W^2_{\rm
 TSC}(\bvec k)$ (upper left)
, its isotropic approximation $W^2_{\rm TSC iso}(\bvec k)$(upper right), 
the difference $W^2_{\rm TSC}(\bvec k)-W^2_{\rm TSC iso}(\bvec k)$ (lower
 left), and the fractional difference $1-W^2_{\rm TSC iso}(\bvec
 k)/W^2_{\rm TSC}(\bvec k)$ (lower right). The wavenumber $k$ is normalized
 by $k_{\rm grid}/\pi$.  }
\label{TSC-vs-exponentialcut}
\vspace*{0.5cm}
\end{figure*}

Let us consider the case in which $N$ 
particles are distributed in a square region. 
The surface number density of the particles is given by
\BE
n_0(\bvec x) = \sum_{i=1}^N \delta_D(\bvec x-\bvec x'), 
\EE
where $x_i$ is the position of $i$-th particle, and
$\delta_D$ is the Dirac delta function.
The smoothed surface number density is given by the convolution
of $n_0$ and the window function $W(x)$,
\BE
n(\bvec x) =\int_V d^2 \bvec x'\, n_0(\bvec x')W(\bvec x-\bvec x'),
\EE
where $W(\bvec x)$ is the window function that specifies the smoothing
scheme. The window function of the TSC scheme is
\BE
W_{\rm TSC}(\bvec{x})=W_{\rm TSC}(x_1)W_{\rm TSC}(x_2),
\EE
where
\BE
  W_{\rm TSC}(x)=\frac{1}{r_{\rm{grid}}} \left \{  
\begin{array}{ll}\!
3/4-(x/{r_{\rm grid}})^2,\,\,\,\, |x|\le \frac{r_{\rm
 grid}}{2} 
\\
\! \!(3/2-|x|/r_{\rm grid})^2/2,\,\,\,\, 
\frac{r_{\rm grid}}{2} \le |x| \le \frac{3 r_{\rm grid}}{2}
\\
\! \! 0,\,\,\,\, \rm{otherwise.}
\end{array}
\right.
\EE
The Fourier transformation is given by 
\BE
W_{\rm TSC}(\bvec{k})={\rm{sinc}}^3 \biggl(\frac{\pi k_1}{k_{\rm grid}}
\biggr)
{\rm{sinc}}^3 \biggl(\frac{\pi k_2}{k_{\rm grid}} \biggr).
\EE
As $\rm{sinc}^6(x)\approx \exp[-x^2]$ for $|x| \ll 1$, we can
approximate the window function by isotropic one
\BE
W^2_{\rm TSC}(\bvec{k})\approx W^2_{\rm TSC iso}(k)=
\exp ( -\pi^2 k^2/ k_{\rm grid}^2 ).
\EE
For simplicity, we use this isotropic approximated
window function instead of the exact anisotropic one. The reason
can be explained as follows.
As shown in Fig \ref{TSC-vs-exponentialcut}, the difference between the original
and isotropic one dimensional squared window functions is $<0.02$ 
and the fractional difference is $\lesssim 0.3$ for $x= \pi k/k_{\rm grid} <0.5$. 
The difference becomes conspicuous at $x>0.5$, which originates 
from the anisotropy of grids. However, as long as the smoothing scheme is 
robust, such differences should not affect the result as the differences
are limited to the tail part.
For instance, in our case, the Poisson noise dominates the signal on
very small scales, which correspond to the tail part. 
Therefore, the tiny error in the tail of the isotropic 
window function does not affect the result as it is almost masked by
the Poisson noise. In terms of $r_{\rm{grid}}=2 \pi/ k_{\rm grid}$, the
isotropic window function can be also written as 
\BE
W^2_{\rm TSC iso}(k)= \exp[-k^2 r^2_{\rm{grid}}/4].
\EE         
\section{Constant-shift cut}
\begin{figure*}
\vspace*{1.0cm}
\includegraphics[width=160mm]{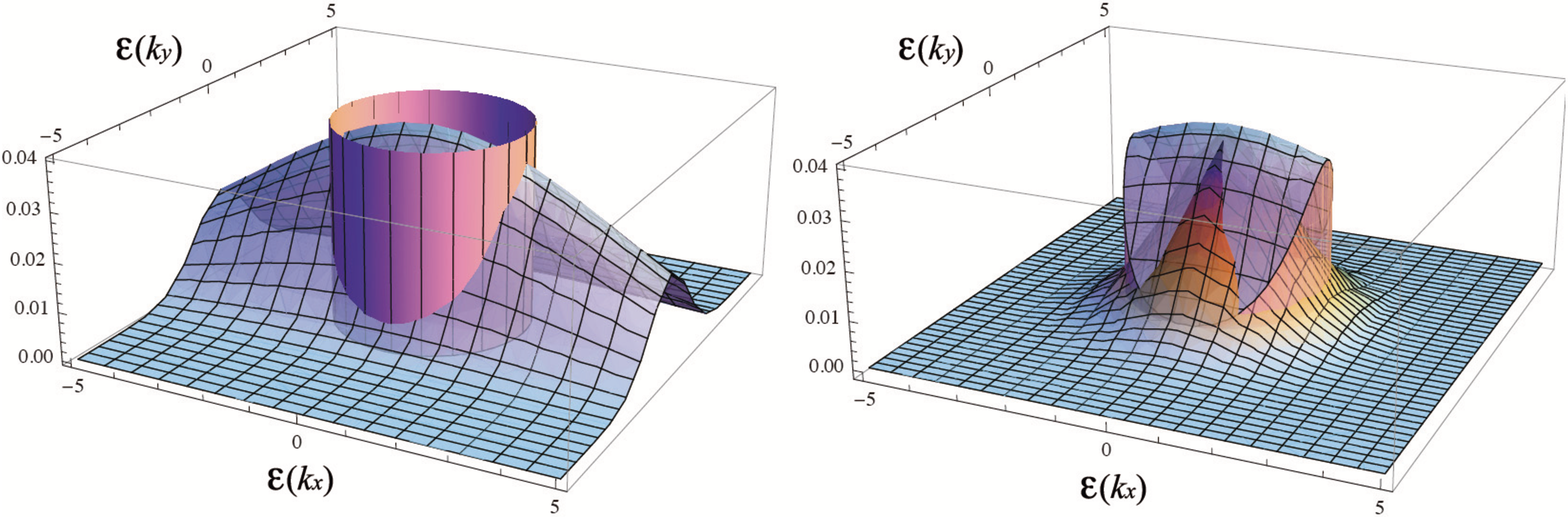}
\caption{Plots of the original Gaussian 
PDF (left) and
the constrained PDF with its Gaussian approximation (right). 
The constrained PDF is obtained by imposing a condition 
$\varepsilon^2(k_x)+  \varepsilon^2(k_y)   \le \varepsilon^2(\textrm{obs})$. Here
we assume $\sigma_x=4$ and $\sigma_y=1$.    }
\label{CScut}
\vspace*{0.5cm}j
\end{figure*}

The use of the constant-shift (CS) cut is verified 
as follows. The variance of the shift $\varepsilon$ 
of a pair of lensed images separated by an angle $\theta_{\max}$
is approximately given by a finite sum
\BE
\langle \varepsilon^2 \rangle =\sum_i 
\frac{d \langle \varepsilon^2(\theta_{\max},k_i) \rangle }{d \ln k} 
\Delta \ln k_i,
\EE
where $\Delta \ln k_i $ is a logarithmic interval in the wavenumber $k$.
Although the correlation between different modes is not negligible
due to the non-linearity of fluctuations at scales $\lesssim
1\,$Mpc, binning of different modes may weaken such a correlation.
Suppose a simplest case in which the variance of the shift is 
approximately given by independent two modes with wavenumbers 
$k_x,k_y (k_x>k_y)$ and the both modes $\varepsilon(k_x), \varepsilon(k_y)$ 
obey a Gaussian distribution
with variances $\sigma_x, \sigma_y (\sigma_x>\sigma_y)$ 
as a result of binning. If the variance of the shift
$\langle \varepsilon^2 \rangle=\langle \varepsilon^2(k_x)+  \varepsilon^2(k_y)
\rangle $ is larger than $\varepsilon^2(\textrm{obs})$, we have
to impose a restriction $ \varepsilon^2(k_x)+  \varepsilon^2(k_y)   \le
\varepsilon^2(\textrm{obs})$ on these modes. In other words, the
domain of the input modes should be restricted on a disk with a
radius $\varepsilon(\textrm{obs})$ (see Fig. \ref{CScut}). 
Consequently, the widths of the constrained PDF are the same
except for modes (such as $k_y$) whose widths of the original PDF are smaller 
than the radius. Approximating these constrained PDFs by Gaussian PDFs
with a fixed variance that corresponds to the size of the radius, one
obtains the constant-shift (CS) cut, which makes the power of the shift
constant at modest scales $k_{\textrm{lens}}<k<k_{\textrm{cut}}$.
In a similar manner, one can easily 
apply these arguments to the cases with many modes.   
In general, the strong mode-coupling will inhibit our
Gaussian approximation. However, the constraint on the shift
naturally leads to a selection of modes with small fluctuations. 
Therefore, we expect that the Gaussian assumption
works well even in the strong non-linear regime
if the constraints on the positions are stringent enough.
\bibliographystyle{mn2e}
\bibliography{weak-lensing-by-los2}

\begin{thebibliography}{}

\bibitem[\protect\citeauthoryear{Amara, Metcalf, Cox \& Ostriker}{Amara
  et~al.}{2006}]{amara2006}
Amara A.,  Metcalf R.~B.,  Cox T.~J.,    Ostriker J.~P.,  2006, Monthly Notices
  of the Royal Astronomical Society, 367, 1367

\bibitem[\protect\citeauthoryear{Chen}{Chen}{2009}]{chen2009}
Chen J.,  2009, Astronomy \& Astrophysics, 498, 49

\bibitem[\protect\citeauthoryear{Chen, Koushiappas \& Zentner}{Chen
  et~al.}{2011}]{chen2011}
Chen J.,  Koushiappas S.~M.,    Zentner A.~R.,  2011, Astrophysical Journal,
  741

\bibitem[\protect\citeauthoryear{Chen, Kravtsov \& Keeton}{Chen
  et~al.}{2003}]{chen2003}
Chen J.,  Kravtsov A.~V.,    Keeton C.~R.,  2003, Astrophysical Journal, 592,
  24

\bibitem[\protect\citeauthoryear{Chiba}{Chiba}{2002}]{chiba2002}
Chiba M.,  2002, The Astrophysical Journal, 565, 17

\bibitem[\protect\citeauthoryear{Chiba, Minezaki, Kashikawa, Kataza \&
  Inoue}{Chiba et~al.}{2005}]{chiba2005}
Chiba M.,  Minezaki T.,  Kashikawa N.,  Kataza H.,    Inoue K.~T.,  2005,
  Astrophysical Journal, 627, 53

\bibitem[\protect\citeauthoryear{{Coles} \& {Jones}}{{Coles} \&
  {Jones}}{1991}]{coles91}
{Coles} P.,  {Jones} B.,  1991, Monthly Notices of the Royal Astronomical
  Society, 248, 1

\bibitem[\protect\citeauthoryear{Crocce, Pueblas \& Scoccimarro}{Crocce
  et~al.}{2006}]{crocce2006}
Crocce M.,  Pueblas S.,    Scoccimarro R.,  2006, Monthly Notices of the Royal
  Astronomical Society, 373, 369

\bibitem[\protect\citeauthoryear{{Das} \& {Ostriker}}{{Das} \&
  {Ostriker}}{2006}]{das2006}
{Das} S.,  {Ostriker} J.~P.,  2006, Astrophysical Journal, 645, 1

\bibitem[\protect\citeauthoryear{Eisenstein \& Hu}{Eisenstein \&
  Hu}{1999}]{eisenstein1999}
Eisenstein D.~J.,  Hu W.,  1999, Astrophysical Journal, 511, 5

\bibitem[\protect\citeauthoryear{{Falco}, {Gorenstein} \& {Shapiro}}{{Falco}
  et~al.}{1985}]{falco85}
{Falco} E.~E.,  {Gorenstein} M.~V.,    {Shapiro} I.~I.,  1985, Astrophysical
  Journal Letter, 289, L1

\bibitem[\protect\citeauthoryear{Goicoechea \& Shalyapin}{Goicoechea \&
  Shalyapin}{2010}]{goicoechea2010}
Goicoechea L.~J.,  Shalyapin V.~N.,  2010, Astrophysical Journal, 708, 995

\bibitem[\protect\citeauthoryear{{Hewitt}, {Turner}, {Lawrence}, {Schneider} \&
  {Brody}}{{Hewitt} et~al.}{1992}]{hewitt92}
{Hewitt} J.~N.,  {Turner} E.~L.,  {Lawrence} C.~R.,  {Schneider} D.~P.,
  {Brody} J.~P.,  1992, Astronomical Journal, 104, 968

\bibitem[\protect\citeauthoryear{{Hilbert}, {White}, {Hartlap} \&
  {Schneider}}{{Hilbert} et~al.}{2007}]{hilbert2007}
{Hilbert} S.,  {White} S.~D.~M.,  {Hartlap} J.,    {Schneider} P.,  2007,
  MNRAS, 382, 121

\bibitem[\protect\citeauthoryear{Huchra, Gorenstein, Kent, Shapiro, Smith,
  Horine \& Perley}{Huchra et~al.}{1985}]{Huchra1985}
Huchra J.,  Gorenstein M.,  Kent S.,  Shapiro I.,  Smith G.,  Horine E.,
  Perley R.,  1985, Astronomical Journal, 90, 691

\bibitem[\protect\citeauthoryear{Inoue \& Chiba}{Inoue \&
  Chiba}{2005a}]{inoue2005a}
Inoue K.~T.,  Chiba M.,  2005a, Astrophysical Journal, 634, 77

\bibitem[\protect\citeauthoryear{Inoue \& Chiba}{Inoue \&
  Chiba}{2005b}]{inoue2005b}
Inoue K.~T.,  Chiba M.,  2005b, Astrophysical Journal, 633, 23

\bibitem[\protect\citeauthoryear{Inoue \& Takahashi}{Inoue \&
  Takahashi}{2012}]{it2012}
Inoue K.~T.,  Takahashi R.,  2012, MNRAS, 426, 2978

\bibitem[\protect\citeauthoryear{{Jain}, {Seljak} \& {White}}{{Jain}
  et~al.}{2000}]{jain2000}
{Jain} B.,  {Seljak} U.,    {White} S.,  2000, Astrophysical Journal, 530, 547

\bibitem[\protect\citeauthoryear{Jarosik, Bennett, Dunkley, Gold, Greason,
  Halpern, Hill, Hinshaw, Kogut, Komatsu, Larson, Limon, Meyer, Nolta, Odegard,
  Page, Smith, Spergel, Tucker, Weiland, Wollack \& Wright}{Jarosik
  et~al.}{2011}]{jarosik2011}
Jarosik N.,  Bennett C.~L.,  Dunkley J.,  Gold B.,  Greason M.~R.,  Halpern M.,
   Hill R.~S.,  Hinshaw G.,  Kogut A.,  Komatsu E.,  Larson D.,  Limon M.,
  Meyer S.~S.,  Nolta M.~R.,  Odegard N.,  Page L.,  Smith K.~M.,  Spergel
  D.~N.,  Tucker G.~S.,  Weiland J.~L.,  Wollack E.,    Wright E.~L.,  2011,
  Astrophysical Journal Supplement Series, 192, 1

\bibitem[\protect\citeauthoryear{{Jing}, {Zhang}, {Lin}, {Gao} \&
  {Springel}}{{Jing} et~al.}{2006}]{jing06}
{Jing} Y.~P.,  {Zhang} P.,  {Lin} W.~P.,  {Gao} L.,    {Springel} V.,  2006,
  Astrophysical Journal Letter, 640, L119

\bibitem[\protect\citeauthoryear{{Kayo}, {Taruya} \& {Suto}}{{Kayo}
  et~al.}{2001}]{kayo01}
{Kayo} I.,  {Taruya} A.,    {Suto} Y.,  2001, Astrophysical Journal, 561, 22

\bibitem[\protect\citeauthoryear{{Kofman}, {Bertschinger}, {Gelb}, {Nusser} \&
  {Dekel}}{{Kofman} et~al.}{1994}]{kofman94}
{Kofman} L.,  {Bertschinger} E.,  {Gelb} J.~M.,  {Nusser} A.,    {Dekel} A.,
  1994, Astrophysical Journal, 420, 44

\bibitem[\protect\citeauthoryear{{Koopmans}}{{Koopmans}}{2005}]{koopmans2005}
{Koopmans} L.~V.~E.,  2005, Monthly Notices of Royal Astronomical Society, 363,
  1136

\bibitem[\protect\citeauthoryear{Kundic, Hogg, Blandford, Cohen, Lubin \&
  Larkin}{Kundic et~al.}{1997}]{kundic1997b}
Kundic T.,  Hogg D.~W.,  Blandford R.~D.,  Cohen J.~G.,  Lubin L.~M.,    Larkin
  J.~E.,  1997, Astronomical Journal, 114, 2276

\bibitem[\protect\citeauthoryear{Lawrence, Elston, Januzzi \& Turner}{Lawrence
  et~al.}{1995}]{lawrence1995}
Lawrence C.~R.,  Elston R.,  Januzzi B.~T.,    Turner E.~L.,  1995,
  Astronomical Journal, 110, 2570

\bibitem[\protect\citeauthoryear{Maccio \& Miranda}{Maccio \&
  Miranda}{2006}]{maccio2006}
Maccio A.~V.,  Miranda M.,  2006, Monthly Notices of the Royal Astronomical
  Society, 368, 599

\bibitem[\protect\citeauthoryear{McKean, Koopmans, Flack, Fassnacht, Thompson,
  Matthews, Blandford, Readhead \& Soifer}{McKean et~al.}{2007}]{mckean2007}
McKean J.~P.,  Koopmans L. V.~E.,  Flack C.~E.,  Fassnacht C.~D.,  Thompson D.,
   Matthews K.,  Blandford R.~D.,  Readhead A. C.~S.,    Soifer B.~T.,  2007,
  Monthly Notices of the Royal Astronomical Society, 378, 109

\bibitem[\protect\citeauthoryear{{MacLeod}, {Jones}, {Agol} \&
  {Kochanek}}{{MacLeod} et~al.}{2012}]{macleod2012}
{MacLeod} C.~L.,  {Jones} R.,  {Agol} E.,    {Kochanek} C.~S.,  2012,
  ArXiv:1212.2166

\bibitem[\protect\citeauthoryear{MacLeod, Kochanek \& Agol}{MacLeod
  et~al.}{2009}]{macleod2009}
MacLeod C.~L.,  Kochanek C.~S.,    Agol E.,  2009, Astrophysical Journal, 703,
  1177

\bibitem[\protect\citeauthoryear{Magain, Surdej, Swings, Borgeest, Kayser,
  Kuhr, Refsdal \& Remy}{Magain et~al.}{1988}]{magain1988}
Magain P.,  Surdej J.,  Swings J.~P.,  Borgeest U.,  Kayser R.,  Kuhr H.,
  Refsdal S.,    Remy M.,  1988, Nature, 334, 325

\bibitem[\protect\citeauthoryear{Mao \& Schneider}{Mao \&
  Schneider}{1998}]{mao1998}
Mao S.,  Schneider P.,  1998, Monthly Notices of the Royal Astronomical
  Society, 295, 587

\bibitem[\protect\citeauthoryear{Metcalf}{Metcalf}{2005}]{metcalf2005a}
Metcalf R.~B.,  2005, The Astrophysical Journal, 629, 673

\bibitem[\protect\citeauthoryear{Metcalf \& Madau}{Metcalf \&
  Madau}{2001}]{metcalf2001}
Metcalf R.~B.,  Madau P.,  2001, The Astrophysical Journal, 563, 9

\bibitem[\protect\citeauthoryear{Metcalf, Moustakas, Bunker \& Parry}{Metcalf
  et~al.}{2004}]{metcalf2004}
Metcalf R.~B.,  Moustakas L.~A.,  Bunker A.~J.,    Parry I.~R.,  2004,
  Astrophysical Journal, 607, 43

\bibitem[\protect\citeauthoryear{Minezaki, Chiba, Kashikawa, Inoue \&
  Kataza}{Minezaki et~al.}{2009}]{minezaki2009}
Minezaki T.,  Chiba M.,  Kashikawa N.,  Inoue K.~T.,    Kataza H.,  2009,
  Astrophysical Journal, 697, 610

\bibitem[\protect\citeauthoryear{Miranda \& Maccio}{Miranda \&
  Maccio}{2007}]{miranda2007}
Miranda M.,  Maccio A.~V.,  2007, Monthly Notices of the Royal Astronomical
  Society, 382, 1225

\bibitem[\protect\citeauthoryear{More, McKean, More, Porcas, Koopmans \&
  Garrett}{More et~al.}{2009}]{more2009}
More A.,  McKean J.~P.,  More S.,  Porcas R.~W.,  Koopmans L. V.~E.,    Garrett
  M.~A.,  2009, Monthly Notices of the Royal Astronomical Society, 394, 174

\bibitem[\protect\citeauthoryear{Nishimichi, Shirata, Taruya, Yahata, Saito,
  Suto, Takahashi, Yoshida, Matsubara, Sugiyama, Kayo, Jing \&
  Yoshikawa}{Nishimichi et~al.}{2009}]{nishimichi2009}
Nishimichi T.,  Shirata A.,  Taruya A.,  Yahata K.,  Saito S.,  Suto Y.,
  Takahashi R.,  Yoshida N.,  Matsubara T.,  Sugiyama N.,  Kayo I.,  Jing
  Y.~P.,    Yoshikawa K.,  2009, Publications of the Astronomical Society of
  Japan, 61, 321

\bibitem[\protect\citeauthoryear{{Patnaik}, {Browne}, {Walsh}, {Chaffee} \&
  {Foltz}}{{Patnaik} et~al.}{1992}]{patnaik1992}
{Patnaik} A.~R.,  {Browne} I.~W.~A.,  {Walsh} D.,  {Chaffee} F.~H.,    {Foltz}
  C.~B.,  1992, Monthly Notices of the Royal Astronomical Society, 259, 1

\bibitem[\protect\citeauthoryear{Peng}{Peng}{2004}]{peng2004}
Peng C.~Y.,  2004, PhD thesis, Steward Observatory, University of Arizona

\bibitem[\protect\citeauthoryear{Riess, Macri, Casertano, Sosey, Lampeitl,
  Ferguson, Filippenko, Jha, Li, Chornock \& Sarkar}{Riess
  et~al.}{2009}]{riess2009}
Riess A.~G.,  Macri L.,  Casertano S.,  Sosey M.,  Lampeitl H.,  Ferguson
  H.~C.,  Filippenko A.~V.,  Jha S.~W.,  Li W.~D.,  Chornock R.,    Sarkar D.,
  2009, Astrophysical Journal, 699, 539

\bibitem[\protect\citeauthoryear{Ros, Guirado, Marcaide, Perez-Torres, Falco,
  Munoz, Alberdi \& Lara}{Ros et~al.}{2000}]{Ros2000}
Ros E.,  Guirado J.~C.,  Marcaide J.~M.,  Perez-Torres M.~A.,  Falco E.~E.,
  Munoz J.~A.,  Alberdi A.,    Lara L.,  2000, Astronomy and Astrophysics, 362,
  845

\bibitem[\protect\citeauthoryear{Schechter \& Moore}{Schechter \&
  Moore}{1993}]{schechter1993}
Schechter P.~L.,  Moore C.~B.,  1993, Astronomical Journal, 105, 1

\bibitem[\protect\citeauthoryear{{Schneider}, {Kochanek} \&
  {Wambsganss}}{{Schneider} et~al.}{2006}]{skw2006}
{Schneider} P.,  {Kochanek} C.,    {Wambsganss} J.,  2006, {Gravitational
  Lensing: Strong, Weak and Micro}.
Springer-Verlag Berlin Heidelberg New York

\bibitem[\protect\citeauthoryear{{Sluse}, {Chantry}, {Magain}, {Courbin} \&
  {Meylan}}{{Sluse} et~al.}{2012}]{sluse2012}
{Sluse} D.,  {Chantry} V.,  {Magain} P.,  {Courbin} F.,    {Meylan} G.,  2012,
  Astronomy and Astrophysics, 538, A99

\bibitem[\protect\citeauthoryear{{Sluse}, {Kishimoto}, {Anguita}, {Wucknitz} \&
  {Wambsganss}}{{Sluse} et~al.}{2013}]{sluse13}
{Sluse} D.,  {Kishimoto} M.,  {Anguita} T.,  {Wucknitz} O.,    {Wambsganss} J.,
   2013, Astronomy \& Astrophysics, 553, A53

\bibitem[\protect\citeauthoryear{Sluse, Surdej, Claeskens, Hutsemekers, Jean,
  Courbin, Nakos, Billeres \& Khmil}{Sluse et~al.}{2003}]{sluse2003}
Sluse D.,  Surdej J.,  Claeskens J.~F.,  Hutsemekers D.,  Jean C.,  Courbin F.,
   Nakos T.,  Billeres M.,    Khmil S.~V.,  2003, Astronomy and Astrophysics,
  406, L43

\bibitem[\protect\citeauthoryear{Smith, Peacock, Jenkins, White, Frenk, Pearce,
  Thomas, Efstathiou \& Couchman}{Smith et~al.}{2003}]{smith2003}
Smith R.~E.,  Peacock J.~A.,  Jenkins A.,  White S. D.~M.,  Frenk C.~S.,
  Pearce F.~R.,  Thomas P.~A.,  Efstathiou G.,    Couchman H. M.~P.,  2003,
  Monthly Notices of the Royal Astronomical Society, 341, 1311

\bibitem[\protect\citeauthoryear{Springel}{Springel}{2005}]{springel2005}
Springel V.,  2005, Monthly Notices of the Royal Astronomical Society, 364,
  1105

\bibitem[\protect\citeauthoryear{Springel, Yoshida \& White}{Springel
  et~al.}{2001}]{springel2001}
Springel V.,  Yoshida N.,    White S. D.~M.,  2001, New Astronomy, 6, 79

\bibitem[\protect\citeauthoryear{{Stalevski}, {Jovanovi{\'c}}, {Popovi{\'c}} \&
  {Baes}}{{Stalevski} et~al.}{2012}]{stalevski12}
{Stalevski} M.,  {Jovanovi{\'c}} P.,  {Popovi{\'c}} L.~{\v C}.,    {Baes} M.,
  2012, Monthly Notices of Royal Astronomical Society, 425, 1576

\bibitem[\protect\citeauthoryear{Sugai, Kawai, Shimono, Hattori, Kosugi,
  Kashikawa, Inoue \& Chiba}{Sugai et~al.}{2007}]{sugai2007}
Sugai H.,  Kawai A.,  Shimono A.,  Hattori T.,  Kosugi G.,  Kashikawa N.,
  Inoue K.~T.,    Chiba M.,  2007, Astrophysical Journal, 660, 1016

\bibitem[\protect\citeauthoryear{{Takahashi}, {Oguri}, {Sato} \&
  {Hamana}}{{Takahashi} et~al.}{2011}]{takahashi2011}
{Takahashi} R.,  {Oguri} M.,  {Sato} M.,    {Hamana} T.,  2011, Astrophysical
  Journal, 742, 15

\bibitem[\protect\citeauthoryear{{Takahashi}, {Sato}, {Nishimichi}, {Taruya} \&
  {Oguri}}{{Takahashi} et~al.}{2012}]{takahashi2012}
{Takahashi} R.,  {Sato} M.,  {Nishimichi} T.,  {Taruya} A.,    {Oguri} M.,
  2012, Astrophysical Journal, 761, 152

\bibitem[\protect\citeauthoryear{{Taruya}, {Takada}, {Hamana}, {Kayo} \&
  {Futamase}}{{Taruya} et~al.}{2002}]{taruya2002}
{Taruya} A.,  {Takada} M.,  {Hamana} T.,  {Kayo} I.,    {Futamase} T.,  2002,
  Astrophysical Journal, 571, 638

\bibitem[\protect\citeauthoryear{Tonry}{Tonry}{1998}]{tonry1998}
Tonry J.~L.,  1998, Astronomical Journal, 115, 1

\bibitem[\protect\citeauthoryear{Tonry \& Kochanek}{Tonry \&
  Kochanek}{1999}]{tonry1999}
Tonry J.~L.,  Kochanek C.~S.,  1999, Astronomical Journal, 117, 2034

\bibitem[\protect\citeauthoryear{{Trotter}, {Winn} \& {Hewitt}}{{Trotter}
  et~al.}{2000}]{trotter2000}
{Trotter} C.~S.,  {Winn} J.~N.,    {Hewitt} J.~N.,  2000, The Astrophysical
  Journal, 535, 671

\bibitem[\protect\citeauthoryear{{van Daalen}, {Schaye}, {Booth} \& {Dalla
  Vecchia}}{{van Daalen} et~al.}{2011}]{vandaalen2011}
{van Daalen} M.~P.,  {Schaye} J.,  {Booth} C.~M.,    {Dalla Vecchia} C.,  2011,
  Monthly Notices of Royal Astronomical Society, 415, 3649

\bibitem[\protect\citeauthoryear{{Weymann}, {Latham}, {Roger}, {Angel},
  {Green}, {Liebert}, {Turnshek}, {Turnshek} \& {Tyson}}{{Weymann}
  et~al.}{1980}]{weymann80}
{Weymann} R.~J.,  {Latham} D.,  {Roger} J.,  {Angel} P.,  {Green} R.~F.,
  {Liebert} J.~W.,  {Turnshek} D.~A.,  {Turnshek} D.~E.,    {Tyson} J.~A.,
  1980, Nature, 285, 641

\bibitem[\protect\citeauthoryear{Wong, Keeton, Williams, Momcheva \&
  Zabludoff}{Wong et~al.}{2011}]{wong2011}
Wong K.~C.,  Keeton C.~R.,  Williams K.~A.,  Momcheva I.~G.,    Zabludoff
  A.~I.,  2011, Astrophysical Journal, 726

\bibitem[\protect\citeauthoryear{Xu, Mao, Wang, Springel, Gao, White, Frenk,
  Jenkins, Li \& Navarro}{Xu et~al.}{2009}]{xu2009}
Xu D.,  Mao S.,  Wang J.,  Springel V.,  Gao L.,  White S.,  Frenk C.,  Jenkins
  A.,  Li G.,    Navarro J.,  2009, Monthly Notices of the Royal Astronomical
  Society, 398, 1235

\bibitem[\protect\citeauthoryear{Xu, Mao, Cooper, Gao, Frenk, Angulo \&
  Helly}{Xu et~al.}{2012}]{xu2012}
Xu D.~D.,  Mao S.,  Cooper A.~P.,  Gao L.,  Frenk C.~S.,  Angulo R.~E.,
  Helly J.,  2012, Monthly Notices of the Royal Astronomical Society, 421, 2553

\bibitem[\protect\citeauthoryear{Xu, Mao, Cooper, Wang, Gao, Frenk \&
  Springel}{Xu et~al.}{2010}]{xu2010}
Xu D.~D.,  Mao S.~D.,  Cooper A.~P.,  Wang J.,  Gao L.~A.,  Frenk C.~S.,
  Springel V.,  2010, Monthly Notices of the Royal Astronomical Society, 408,
  1721

\end{thebibliography}

\end{document}